\documentclass[apj]{emulateapj-rtx4}
\usepackage{times}
\usepackage{mathptmx}
\usepackage{lscape}
\usepackage{xcolor}
\usepackage{lscape}
\slugcomment{Accepted by ApJ for publication}

\begin{document}

\title{NLTE Analysis of High Resolution {\it{H}}-band Spectra. \\I. Neutral Silicon\altaffilmark{$\ast$}}
\altaffiltext{$\ast$}{Based on observations collected on the 2.16m telescope at Xinglong station, National Astronomical Observatories, Chinese Academy of Sciences, the 2.2m telescope at the Calar Alto Observatory, the 1.88m reflector on the Okayama Astrophysical Observatory, the Kitt Peak coud\'e feed telescope, and the McMath-Pierce solar telescope and the coud\'e focus of the Mayall 4m reflector at Kitt Peak.}

\author{Junbo Zhang\altaffilmark{1,2}, Jianrong Shi\altaffilmark{1,2}, Kaike Pan\altaffilmark{3}, Carlos Allende Prieto\altaffilmark{4,5}, Chao Liu\altaffilmark{1}}
\affil{\altaffilmark{1} Key Laboratory of Optical Astronomy, National Astronomical Observatories, Chinese Academy of Sciences, A20 Datun Road, Chaoyang District, Beijing 100012, China}
\affil{\altaffilmark{2} School of Astronomy and Space Science, University of Chinese Academy of Sciences, Beijing 100049, China}
\affil{\altaffilmark{3} Apache Point Observatory and New Mexico State University, P.O. Box 59, Sunspot, NM, 88349-0059, USA}
\affil{\altaffilmark{4} Instituto de Astrof\'isica de Canarias, 38205 La Laguna, Tenerife, Spain}
\affil{\altaffilmark{5} Departamento de Astrof\'isica, Universidad de La Laguna, 38206 La Laguna, Tenerife, Spain}
\email{sjr@bao.ac.cn}

\begin{abstract}
We investigated the reliability of our silicon atomic model and the influence of non-local thermodynamical equilibrium (NLTE) on the formation of neutral silicon (\ion{Si}{1}) lines in the near-infrared (near-IR) {\it{H}}-band. We derived the differential Si abundances for 13 sample stars with high-resolution {\it{H}}-band spectra from the Apache Point Observatory Galactic Evolution Experiment (APOGEE), as well as from optical spectra, both under local thermodynamical equilibrium (LTE) and NLTE conditions. We found that the differences between the Si abundances derived from the {\it{H}}-band and from optical lines for the same stars are less than 0.1\,dex when the NLTE effects included, and that NLTE reduces the line-to-line scatter in the {\it{H}}-band spectra for most sample stars. These results suggest that our Si atomic model is appropriate for studying the formation of {\it{H}}-band Si lines. Our calculations show that the NLTE corrections of the \ion{Si}{1} {\it{H}}-band lines are negative, i.e. the final Si abundances will be overestimated in LTE. The corrections for strong lines depend on surface gravity, and tend to be larger for giants, reaching $\sim -$0.2\,dex in our sample, and up to $\sim -$0.4\,dex in extreme cases of APOGEE targets. Thus, the NLTE effects should be included in deriving silicon abundances from {\it{H}}-band \ion{Si}{1} lines, especially for the cases where only strong lines are available.
\end{abstract}


\section{INTRODUCTION}
Silicon is an important $\alpha$-element mainly produced during oxygen and neon burning, and returned to the interstellar medium by Type-II Supernovae \citep[SNe II;][]{woo95}. SNe Ia may also produce small fraction \citep{tsu95}. Silicon is an important element of the interstellar dust, one of the main electron contributors (only next to Fe and Mg) in the atmospheres of late-type stars \citep{hol73,wed01}. Silicon abundance is often used as a tracer to explore the formation and evolution of the Solar System \citep{joh11,zam13}; and to study the Galactic structure, chemical enrichment history and the origin of the Galaxy in many studies. For example, the silicon abundance, combined with other $\alpha$-elements, is often adopted as an indicator to distinguish stars from different populations, namely thick- and thin-disks \citep{lee11}. A series of studies, e.g. \citet{ben05,ben14,red03,red06,nis10,zha11} have observed amounts of high-resolution spectra and have derived accurate silicon abundances. Comparing with different Galactic evolution models, e.g. \citet{sam98,gos00,rom10,kob11}, these abundances can help astronomers to understand the chemical enrichment history and the origin of the Galaxy. Thus, an accurate measurement of silicon abundances is necessary for many astrophysical applications.

\citet{kam73,kam78,kam82} calculated, both in LTE and NLTE, theoretical equivalent widths and profiles for silicon lines, and compared them with observational data of a dozen early-type stars. The results indicate that the NLTE calculations provided better agreement with observations. The deviations from LTE on Si abundances in the photospheres of the \object{Sun} and \object{Vega} have been investigated by \citet{wed01}, who found that the mean NLTE correction for Si was $\sim -$0.01\,dex for the Sun, and $\sim -$0.054\,dex for \object{Vega}. This indicated that the NLTE effects on the Si abundance in the Sun could be neglected, which was confirmed by \citet{shi08}. Later on, \citet{shi09,shi11} systematically investigated the NLTE effects on the derived silicon abundances in the atmospheres of metal-poor stars based on visible lines, and found the NLTE effects are large for the two strong UV lines at 3905 and 4103\,\AA, especially for warm metal-poor stars. \citet{shi12} extended the study to the near-IR {\it{J}}-band Si lines, and found that the NLTE effects depend on surface gravities becoming larger for giants. Recently, \citet{ber13} investigated the NLTE effects on the {\it{J}}-band Si lines for red supergiants, and confirmed that Si abundance based on NLTE is significantly lower than that from LTE.

Until very recent, almost all observed high-resolution spectra are from UV, optical, and near-IR {\it{J}} bands, therefore, previous studies on NLTE Si abundance are for spectral lines in these three bands. The situation has changed since the Apache Point Observatory Galactic Evolution Experiment (APOGEE) survey\footnote{http://www.sdss.org/surveys/apogee} \citep{maj15} \citep[as part of SDSS-III,][]{eis11} started to take high resolution IR {\it{H}}-band spectra for several hundred thousands stars. Thus, it is highly desirable to extend the NLTE investigations to the {\it{H}}-band where there are a dozen \ion{Si}{1} lines, which are clearly seen in APOGEE spectra.

Since 2011, the APOGEE survey employs a fiber spectrograph that simultaneously records 300 spectra in the {\it{H}}-band between 1.51 and 1.70 $\mu m$ at a spectral resolution of $R \sim$ 22,500. Detailed information about the instrument was provided by \citet{wil10}. Taking advantage of the reduced effect of extinction in IR {\it{H}}-band, APOGEE has observed $\sim$ 150,000 stars, predominantly red giants in all major Galactic components accessible from Apache Point Observatory (APO) \citep{hol15,maj15}. The spectra have been included in the SDSS Data Release 10 (DR10) \citep{ahn14} and SDSS Data Release 12 (DR12) \citep{ala15}. The data provide a promising way to trace and explore the formation history of the Galaxy, and they are revolutionizing our knowledge on kinematical and chemical enrichment history of all Galactic stellar populations.

The APOGEE Stellar Parameters and Chemical Abundances Pipeline (ASPCAP) provides the physical and chemical parameters for the APOGEE stars \citep{gar16}. In addition to the stellar parameters, i.e. the effective temperature ($T_{\rm{eff}}$), surface gravity (log\,$g$), metallicity ([M/H]), ASPCAP delivers individual chemical abundances for 15 elements. The accuracy of these derived stellar fundamental parameters and chemical compositions may be compromised. NLTE effects are enhanced by the characteristic low densities, found in the atmospheres of giants, and the absolute reduction in collision rates, thus, affects the atomic populations \citep{mes13}. \citet{haw16} have performed an independent procedure to determine the chemical abundances of the APOGEE $+$ Kepler stellar sample (APOKASC) and they inferred that the discrepant phenomenon for some elements is likely due to the NLTE effects. As part of a series of studies on NLTE analysis of {\it{H}}-band lines for several important elements, e.g. Na, Mg, Al, Si, K, Ca, Fe, et al., this work aims to validate the Si atomic model, and to investigate how the abundances derived from  the Si {\it{H}}-band transitions are affected by departures from LTE.

This paper is organized as follows. In Section \ref{method}, we briefly introduce our adopted Si model atom and NLTE calculation codes, while the selection of the sample stars and the observed spectra are described in Section \ref{sample_access}. The stellar parameters of our sample stars are determined in Section \ref{para}, and the Si abundances derived from both the {\it{H}}-band and optical lines for the sample stars under LTE and NLTE analysis are presented in Section \ref{nlte}. Finally, the conclusions are given in Section \ref{con}.

\section{METHOD OF NLTE CALCULATIONS} \label{method}
\subsection{Model Atom of Silicon} \label{atom}
\begin{figure*}
\centering
\includegraphics[scale=0.7]{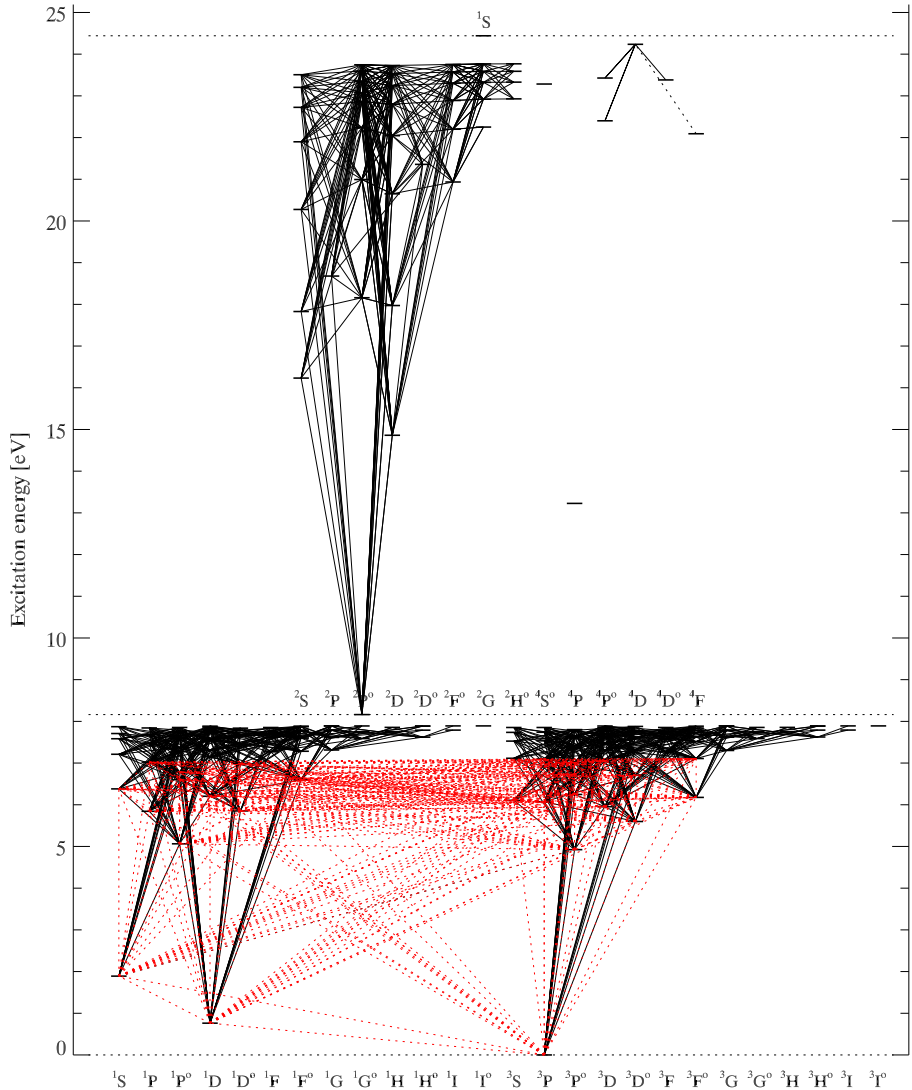}\\
\caption{Grotrian diagram of the silicon model atom. Si II quartets are neglected. Allowed transitions are black continuous, the forbidden Si I 4103\AA\ line is black dotted. Transitions of \ion{Si}{1} between 26 low-lying energy terms presented by \citet{bel14} are shown with red-dotted lines. \label{fig0}}
\end{figure*}

The Si atomic model that we used here is similar to that of \citet{shi08}, which includes the most important 132 terms of \ion{Si}{1}, 41 terms of \ion{Si}{2}, and the \ion{Si}{3} ground state. The radiative data are taken from \citet{nah93}. Lacking accurate values for incollisions with neutral hydrogen, \citet{shi08} suggested S$_{\rm H} = 0.1$ by fitting solar strong infrared \ion{Si}{1} lines. Fortunately, \citet{bel14} calculated the cross sections and rate coefficients for inelastic processes in Si + H and Si$^{\rm+}$ + H$^{\rm-}$ collisions for all transitions between 26 low-lying states plus the ionic state. We revised the Si atomic model by including all cross sections from \citet{bel14} whenever available; Otherwise, S$_{\rm H} =$ 0.1 was adopted. The Grotrian diagram of the silicon model atom with the transitions between 26 low-lying energy terms relative to \citet{bel14} is shown in Fig.\,\ref{fig0}. In this work, we also calculated the NLTE line profiles for the Sun and \object{Arcturus} with the four different collision treatments, i.e. the Drawin recipe with S$_{\rm{H}} =$ 0.0, 0.1 and 1.0 and the treatment from \citet{bel14} and the results are depicted in Fig.\,\ref{fig2} and Fig.\,\ref{fig3} respectively. Our adopted stellar parameters for the Sun are $T_{\rm{eff}}$ $=$ 5777K, [Fe/H] $=$ 0.0\,dex, log\,$g$ $=$ 4.44\,dex and $\xi_t$ $=$ 0.9\,km\,s$^{-1}$. As shown in these two groups of figures, the calculated NLTE line profiles with S$_{\rm H} =$ 0.1 and the treatment from \citet{bel14} can fit the observed spectral lines well for both the Sun and Arcturus; for the strong lines at 15888 and 16680\,\AA, the synthetic line profiles with S$_{\rm H} =$ 1.0 are shallower than the observed ones, while those with S$_{\rm H} =$ 0.0 are slightly deep, with the same silicon abundance.

\subsection{Model Atmospheres} \label{model}
We adopted MARCS atmospheric models\footnote{http://marcs.astro.uu.se}, described in detail by \citet{gus08}. The MARCS models come in two types, the plane-parallel and spherically-symmetric model atmospheres. The models with low surface gravities (-1.0 $\leq$ log\,$g$ $\leq$ 3.5) are calculated in spherical geometry, while the plane-parallel ones are adopted for stars with 3.0 $\leq$ log\,$g \leq 5.5$. \citet{gus08} suggested that sphericity effects are generally important for the temperature structures of low-gravity stars. In this paper, spherical model atmospheres are used for stars with log\,$g$ $\le$ 3.5 and plane-parallel model atmospheres for the others, and they are interpolated with a FORTRAN-based routine coded by Thomas Masseron\footnote{http://marcs.astro.uu.se/software.php}.

The main characteristics of the MARCS model atmospheres are summarized below \citep{gus08}:
\begin{itemize}
\item The basic chemical composition of the Sun in model atmospheres is that listed by \citet{gre07}. The adopted solar Si abundance is 7.51\,dex, which is the 3D-based LTE Si abundance and also the meteoritic Si abundance. \citet{ama16} recently based on a 3D NLTE calculations also found the same solar Si abundance.
\item The $\alpha$-enhancement is included.
\begin{displaymath}
[\alpha/\textrm{Fe}]  =
\left\{
\begin{array}{ll}
0.4 \cdot \left|\textrm{[Fe/H]}\right| & \textrm{if $-$1.0 $<$ [Fe/H] $<$ 0.0}\\
0.4 & \textrm{if [Fe/H] $\le -$1.0}
\end{array} \right.
\end{displaymath}
\item The adopted mixing-length parameter $l/\rm{H}_P$ is 1.5 \citep{hen65}.
\end{itemize}

\subsection{Statistical Equilibrium Codes} \label{codes}
A revised version of DETAIL program \cite[]{but85} was adopted to solve the coupled statistical equilibrium and the radiative transfer equations. This program is based on an accelerated lambda iteration scheme, following the approach described by \citet{ryb91,ryb92}. In this paper, departure coefficients were computed with DETAIL, and then fed to the spectrum synthesis software package Spectrum Investigation Utility (SIU), developed by \citet{ree91} to derive chemical abundances.

\section{THE SAMPLE STARS AND THEIR SPECTRA} \label{sample_access}
\subsection{Sample Selection } \label{sample}
Although \citet{shi12} have demonstrated that their Si atomic model could provide consistent silicon abundances for the optical and infrared {\it{J}}-band spectra, we would like to check whether the atomic model can also be applied to the {\it{H}}-band \ion{Si}{1} lines. We selected 13 FGK dwarfs and giants as sample stars for this test according to the following criteria: 1) they must have available high-resolution (R $>$ 20,000) and high signal-to-noise ratio (S/N $>$ 100) spectra both in the optical and {\it{H}} band; 2) the selected stars should be representative of the typical stellar parameter range of the FGK stars (T$_{eff}$ $\sim$ 4000 $-$ 6500\,K, log\,$g$ $\sim$ 0.0 $-$ 5.0, and [M/H] $\sim$ $-$2.0 $-$ 0.5\,dex). The final stellar parameters of sample stars span from 4275 to 6070\,K for $T_{\rm{eff}}$, from 1.67 to 4.65 for log\,$g$, and from -1.35 to 0.28\,dex for [Fe/H]. However, there are no very metal-poor stars ([Fe/H] $<$ -1.5\,dex) in our sample, due to the weakness of Si lines in the APOGEE spectra for such stars. The IR and optical data are described in the following subsections.

\subsection{Infrared {\it{H}}-band Spectra} \label{H_sample}
The IR {\it{H}}-band spectra of our 13 sample stars are from the NMSU 1m$+$APOGEE observations, and they are included in SDSS DR12. The 1m$+$APOGEE configuration is designed to observe nearby bright stars and to provide an improved calibration  for the main APOGEE survey \citep{hol15}. A bundle of ten fibers was installed connecting the APOGEE instrument to the NMSU (New Mexico State University) 1m telescope. This configuration provides one science fiber and nine sky fibers per observation. Bright stars with magnitude of $0 < H < 8$ are observed in this configuration in dark time when the APOGEE instrument is not connected with the Sloan 2.5m Telescope. The spectra taken with the NMSU 1m$+$APOGEE are reduced and analyzed with the same software employed by the main survey \citep{nid15}. We refer the reader to \citet{feu16} for more details. Since all 13 selected stars are bright, the S/N of APOGEE spectra of these stars are very high (e.g. S/N $\ge$ 400 for \object{Arcturus}). As mentioned earlier, the resolution is about 22,500.
 
The high-resolution (R $\sim$ 500,000) solar infrared spectrum from Kurucz website\footnote{http://kurucz.harvard.edu/sun/irradiance2008/} was adopted in this study. It was obtained by James Brault at Kitt Peak and reduced by Robert L. Kurucz. We also employed the spectrum of \object{Arcturus} from the NOAO science archives\footnote{http://ast.noao.edu/data/other}, which was recorded with the Fourier transform spectrometer \citep[FTS,][]{hal79} operated at the coud\'e focus of the Mayall 4m reflector at Kitt Peak. The detailed description of the observation was presented by \citet{hin95}. The high-resolution ($\sim$ 100,000) and high signal-to-noise ratio spectrum of \object{Arcturus} makes it easier to identify the continuum, and most efficient to recognize blending lines.

\subsection{Optical Spectra} \label{op_sample}
We adopted the optical solar spectrum of \citet{kur84}. Six of our sample stars (\object{HD\,6582, HD\,6920, HD\,102870, HD\,103095, HD\,121370, HD\,148816}) were observed with the fiber-coupled Cassegrain \'echelle spectrograph \citep[FOCES;][]{pfe98} on the 2.2m telescope at the Calar Alto Observatory. Spectra of \object{HD\,87, HD\,22675, HD\,58367, and HD\,177249} were taken with the High Dispersion \'{E}chelle Spectrograph (HIDES) on the coud\'e focus of the 1.88m reflector at the Okayama Astrophysical Observatory \citep{hid03}. The optical spectrum of \object{Arcturus} was obtained with the \'echelle spectrograph (ES) on the Kitt Peak coud\'e feed telescope (KPCFT), with a typical resolving power of 150,000, and a S/N of about 1,000 \citep{hin00, hin05}. Both HD 31501 and HD 67447 were observed using the 2.16m telescope at Xinglong station but with different spectrographs, for \object{HD\,31501} with the Coud\'e \'Echelle Spectrograph\citep[CES;][]{zha01} and for \object{HD\,67447} with the fiber optics \'echelle spectrograph (FOES). The detailed observational information for the sample stars is listed in Table\,\ref{tbl-1} (except \object{the Sun}). It is worthwhile noting that all optical spectra have a resolving power better than 35,000 and S/N $\ge$ 150.

\begin{deluxetable*}{lrllccc}
\tabletypesize{\scriptsize}
\tablecaption{Characteristics of the observed optical spectra\label{tbl-1}}
\tablewidth{0pt}
\tablehead{
\colhead{Star} & \colhead{$V_{mag}$\tablenotemark{1} (mag)} & \colhead{Telescope/spectrograph} & \colhead{Observing run, observer} & \colhead{Spectral range (\AA)} & \colhead{$R$} & \colhead{S/N}
}
\startdata
Arcturus & $-$0.05 & KPCFT/ES & 1998-99, Hinkle K. et al. & 3727-9300 & 150,000 & $\sim$ 1,000\\
HD 87 & 5.55 & 1.88-m/HIDES & Jul. 2007, Anonymous\tablenotemark{2} & 5000-6200 & 67,000 & $\ge$150\\
HD 6582 & 5.17 & 2.2-m/FOCES & Sep. 1995, Fuhrmann K. & 4000-7000 & 35,000 & $\ge$150\\
HD 6920 & 5.67 & 2.2-m/FOCES & Feb. 1997, Fuhrmann K. & 4000-9000 & 60,000 & $\sim$ 200\\
HD 22675 & 5.86 & 1.88-m/HIDES & Jan. 2010, Sato B. & 4000-7540 & 67,000 & $\sim$300\\
HD 31501 & 8.15 & 2.16-m/CES & Jan. 2008,Shi J.R. & 5600-8800 & 40,000 & $\ge$150\\
HD 58367 & 4.99 & 1.88-m/HIDES & Feb. 2004, Anonymous\tablenotemark{2} & 5000-6200 & 67,000 & $\ge$150\\
HD 67447 & 5.34 & 2.16-m/FOES & Jan. 2015, Zhang J.B. & 3900-7260 & 50,000 & $\ge$150\\
HD 102870 & 3.59 & 2.2-m/FOCES & May. 1997, Fuhrmann K. & 4000-9000 & 60,000 & $\sim$ 200\\
HD 103095 & 6.42 & 2.2-m/FOCES & May. 2000, Fuhrmann K. & 4000-9000 & 60,000 & $\sim$ 200\\
HD 121370 & 2.68 & 2.2-m/FOCES & Dec. 1998, Fuhrmann K. & 4000-9000 & 60,000 & $\sim$ 200\\
HD 148816 & 7.27 & 2.2-m/FOCES & Aug. 2001, Gehren T.& 4000-9000 & 60,000 & $\sim$ 200\\
HD 177249 & 5.51 & 1.88-m/HIDES & Nov. 2004, Anonymous\tablenotemark{2} & 5000-6200 & 67,000 & $\sim$300
\enddata
\tablenotetext{1}{Visual magnitudes are derived from the Hipparcos Main Catalogue (ESA 1997) through VizieR\footnote{http://vizier.u-strasbg.fr/viz-bin/VizieR.}.}
\tablenotetext{2}{Spectra were provided by Takeda Y., Sato, B. and Liu Y.J. et al. The observer written in the spectra header is anonymous, and it is difficult for us to identify the actual observers.}
\end{deluxetable*}

\section{STELLAR PARAMETERS} \label{para}
The stellar parameters of all 13 stars were determined via the spectroscopic approach. Specifically, the effective temperature and surface gravity were determined by fulfilling the excitation equilibrium of \ion{Fe}{1} and the ionization equilibrium of \ion{Fe}{1} and \ion{Fe}{2}, respectively; the micro-turbulence velocity was determined by forcing [Fe/H] from different \ion{Fe}{1} lines to be independent of their equivalent widths. Table\,\ref{tbl-8} gives the equivalent widths for our sample stars.

This process of determining stellar parameters is an iterative procedure.  A set of initial parameters is needed to begin with. The initial temperature was derived from the Balmer lines (H$_{\alpha}$ and H$_{\beta}$) \citep{fuh98} if these lines were available. Otherwise, it was obtained based on the color index ($b-y$ or $V-K$) employing the calibration given by \citet{alo96, alo99, alo01}; The initial surface gravity was estimated using the parallax method. There are 30 Fe I and six Fe II optical lines included in our analysis. The line data as well as the equivalent widths for the solar iron lines are listed in Table\,\ref{tbl-6}. Departures from LTE have been considered when determining the iron abundance based on the iron model atom from \citet{mas11} and for the Sun and our sample stars, they are small, less than 0.05\,dex. In Table\,\ref{tbl-6}, we also present the solar LTE and NLTE iron abundances based on the oscillator strengths (log\,$gf$) values recommended by the VALD3 database\footnote{http://vald.astro.uu.se}. According to this table, the iron abundances derived from \ion{Fe}{1} lines are 7.56\,$\pm$\,0.13\,dex in LTE and 7.60\,$\pm$\,0.13\,dex in NLTE, while they are 7.49\,$\pm$\,0.04\,dex from \ion{Fe}{2} lines in both cases. The statistical error for the Sun is uncomfortably large, up to 0.13\,dex, thus we derived the empirical log\,$gf$ by fitting the solar spectrum and presented the NLTE one in this table. The values of log\,$C_{6}$ were calculated referring to \citet{ans91,ans95} and \citet{bar00}.  Based on multiple iterative processes, we estimated the typical uncertainties of $T_{\rm{eff}}$, log\,$g$, [Fe/H], and $\xi_t$ are $\pm$80\,K, $\pm$0.1\,dex, $\pm$0.08\,dex, and 0.2\,km\,s$^{-1}$ respectively.

The final derived stellar parameters, along with stellar parameters for the same stars from the literature, are presented in Table\,\ref{tbl-2}. Our derived values are consistent with those from literature, except log\,$g$ for \object{HD\,22675} and \object{HD\,177249}. Our newly derived log\,$g$ values for the two stars are 0.26 and 0.11, respectively, higher than those determined by \citet{tak08}. We note that our spectroscopic log\,$g$ values agree well with those from the parallax method derived by \citet{tak08}. This may indicate that our spectroscopic surface gravities for these two stars are better than those from \citet{tak08}.

\begin{deluxetable}{lclrcl}
\tabletypesize{\scriptsize}
\tablecaption{Comparison of stellar parameters with other studies\label{tbl-2}}
\tablewidth{0pt}
\tablehead{
\colhead{Star} & \colhead{$T_{\rm{eff}}$} & \colhead{log\,$g$} & \colhead{[Fe/H]} & \colhead{$\xi_t$} & \colhead{Ref.\tablenotemark{a}}\\
               &             (K)          &  [cgs]            &                  &     km\,s$^{-1}$   & \\
}
\startdata
Arcturus               & 4275 & 1.67 & $-$0.58 & 1.60 & This study\\
                       & 4286 & 1.66 & $-$0.52 & 1.74 & RAM11\\
                       & 4286 & 1.66 & $-$0.48 & 1.74 & SHE15\\
HD\,87                 & 5053 & 2.71 & $-$0.10 & 1.35 & This study\\
                       & 5072 & 2.63 & $-$0.10 & 1.35 & TAK08\\
HD\,6582               & 5390 & 4.42 & $-$0.81 & 0.90 & This study\\
                       & 5387 & 4.45 & $-$0.83 & 0.89 & FUH98\\
HD\,6920               & 5845 & 3.45 & $-$0.06 & 1.40 & This study\\
                       & 5838 & 3.48 & $-$0.05 & 1.35 & FUH98\\
\underline{HD\,22675}  & 4901 & 2.76 & $-$0.05 & 1.30 & This study\\
                       & 4878 & 2.50\tablenotemark{b} & $-$0.06 & 1.29 & TAK08\\
                       &      & 2.66\tablenotemark{c} &         &      & TAK08\\
HD\,31501              & 5320 & 4.45 & $-$0.40 & 1.00 & This study\\
                       & 5326 & 4.41 & $-$0.38 & 1.00 & WAN09\\
HD\,58367              & 4932 & 1.79 & $-$0.18 & 2.00 & This study\\
                       & 4911 & 1.76 & $-$0.14 & 2.04 & TAK08   \\
HD\,67447              & 4933 & 2.17 & $-$0.05 & 2.12 & This study\\
                       & 4974 & 2.12 & $-$0.06 & 2.12 & TAK08    \\
HD\,102870             & 6070 & 4.08 &    0.20 & 1.20 & This study \\
                       & 6085 & 4.04 &    0.14 & 1.38 & FUH98\\
                       & 6060 & 4.11 &    0.18 & 1.20 & MAS11\\
HD\,103095             & 5085 & 4.65 & $-$1.35 & 0.80 & This study\\
                       & 5110 & 4.66 & $-$1.35 & 0.85 & FUH98\\
                       & 5070 & 4.69 & $-$1.35 & 0.80 & MAS07\\
HD\,121370             & 6020 & 3.80 &    0.28 & 1.40 & This study \\
                       & 6023 & 3.76 &    0.28 & 1.40 & FUH98\\
HD\,148816             & 5830 & 4.10 & $-$0.73 & 1.40 & This study \\
                       & 5823 & 4.13 & $-$0.73 & 1.40 & NIS10 \\
\underline{HD\,177249} & 5273 & 2.66 &    0.03 & 1.65 & This study\\
                       & 5251 & 2.55\tablenotemark{b} &  0.00 & 1.65 & TAK08 \\
                       &      & 2.62\tablenotemark{c} &       &      & TAK08
\enddata
\tablenotetext{a}{RAM11: \citet{ram11}, SHE15: \citet{she15}, TAK08: \citet{tak08}, FUH98: \citet{fuh98}, WAN09: \citet{wan09}, MAS11: \citet{mas11}, MAS07: \citet{mas07}, NIS10: \citet{nis10}}
\tablenotetext{b}{log\,$g$ derived from the spectroscopic method}
\tablenotetext{c}{log\,$g$ derived from the parallax and evolution-track method}
\tablecomments{Underlines mean these stars with them are discussed in detail in Section \ref{para}.}
\end{deluxetable}

\section{NLTE CALCULATIONS FOR SAMPLE STARS} \label{nlte}
\subsection{Line Data} \label{line}
\subsubsection{Infrared Atomic Line Data in the H band} \label{h_line}
Initially, we found 11 \ion{Si}{1} lines in the {\it{H}}-band APOGEE spectra. A further investigation reveals that seven of them are very weak or heavily blended. As a result, only four lines were employed in this study. The details about them are presented in Table\,\ref{tbl-3}. The transitions are taken from the NIST database\footnote{http://physics.nist.gov/}. The van der Waals damping constants (log\,$C_6$) are extracted from \citet{mel99} , who calculated $C_6$ based on the quantum-mechanical approximate cross sections provided by \citet{ans95}, \citet{bar97} and \cite{bar98}. We derived the solar LTE and NLTE Si abundances using the $gf$-values referring to the references and found that the statistic errors are very large in both LTE and NLTE. In order to reduced the importance of oscillator strengths, therefore, we performed a line-to-line differential analysis and the $gf$-values derived from the LTE and NLTE solar spectrum fits are also listed in Table\,\ref{tbl-3}.

\begin{deluxetable*}{lccrcccccrr}
\tabletypesize{\scriptsize}
\tablecaption{Atomic data of the silicon optical and {\it{H}}-band lines, the derived LTE and NLTE solar silicon abundance based on log\,$gf$ from references and the NLTE corrections for the solar silicon lines \label{tbl-3}}
\tablewidth{0pt}
\tablehead{
\colhead{$\lambda$} & \colhead{Transition} & \colhead{$\chi$} & \colhead{log\,$C_6$} & \colhead{log\,$gf$} & \colhead{Ref.} & \colhead{$\rm{log}\,\varepsilon_{\sun}Si$} & \colhead{$\rm{log}\,\varepsilon_{\sun}Si$} & \colhead{log\,$gf'$} & \colhead{log\,$gf'$} & \colhead{$\Delta_{\sun}$}\\
\colhead{} & \colhead{} & \colhead{} & \colhead{} & \colhead{} & \colhead{} & \colhead{LTE} & \colhead{NLTE} & \colhead{LTE} & \colhead{NLTE} & \colhead{}\\
\colhead{(\AA)} & \colhead{} & \colhead{(eV)} & \colhead{} & \colhead{} & \colhead{} & \colhead{(dex)} & \colhead{(dex)} & \colhead{} & \colhead{} & \colhead{(dex)}
}
\startdata
5701.104 & 4$s~^{\rm{3}}$P$^{\rm{o}}_{\rm{1}}$$-$5$p~^{\rm{3}}$P$_{\rm{0}}$ & 4.930 & $-$30.094 & $-$2.05 & GAR73,KEL08 & 7.60 & 7.60 & $-$1.96 & $-$1.96 & 0.00\\
5772.146 & 4$s~^{\rm{1}}$P$^{\rm{o}}_{\rm{1}}$$-$5$p~^{\rm{1}}$S$_{\rm{0}}$ & 5.082 & $-$30.087 & $-$1.75 & GAR73,KEL08 & 7.64 & 7.63 & $-$1.62 & $-$1.63 & $-0.01$\\
6142.483 & 3$p^{\rm{3}}~^{\rm{3}}$D$^{\rm{o}}_{\rm{1}}$$-$5$f~^{\rm{3}}$D$_{\rm{3}}$ & 5.619 & $-$29.669 & $-1.30$ & KUR07 & 7.37 & 7.37 & $-$1.44 & $-$1.44 & 0.00\\
6145.016 & 3$p^{\rm{3}}~^{\rm{3}}$D$^{\rm{o}}_{\rm{2}}$$-$5$f~^{\rm{3}}$G$_{\rm{3}}$ & 5.616 & $-$29.669 & $-1.31$ & KUR07 & 7.45 & 7.45 & $-$1.37 & $-$1.37 & 0.00\\
6155.134 & 3$p^{\rm{3}}~^{\rm{3}}$D$^{\rm{o}}_{\rm{3}}$$-$5$f~^{\rm{3}}$G$_{\rm{4}}$ & 5.619 & $-$29.669 & $-0.76$ & KUR07 & 7.50 & 7.49 & $-$0.77 & $-$0.78 & $-$0.01\\
6237.319 & 3$p^{\rm{3}}~^{\rm{3}}$D$^{\rm{o}}_{\rm{1}}$$-$5$f~^{\rm{3}}$F$_{\rm{2}}$ & 5.614 & $-$29.669 & $-0.98$ & KUR07 & 7.43 & 7.43 & $-$1.06 & $-$1.06 & 0.00\\
6243.815 & 3$p^{\rm{3}}~^{\rm{3}}$D$^{\rm{o}}_{\rm{2}}$$-$5$f~^{\rm{3}}$F$_{\rm{3}}$ & 5.616 & $-$29.669 & $-1.24$ & KUR07 & 7.49 & 7.49 & $-$1.26 & $-$1.26 & 0.00\\
6244.466 & 3$p^{\rm{3}}~^{\rm{3}}$D$^{\rm{o}}_{\rm{2}}$$-$5$f~^{\rm{1}}$D$_{\rm{2}}$ & 5.616 & $-$29.669 & $-1.09$ & KUR07 & 7.35 & 7.35 & $-$1.25 & $-$1.25 & 0.00\\
mean     &                                                                           &       &           &         &       & 7.48 & 7.48 &         &         &     \\
$\sigma$ &                                                                           &       &           &         &       & 0.10 & 0.10 &         &         &     \\
\hline
15888.440 & 4$s~^{\rm{1}}$P$^{\rm{o}}_{\rm{1}}$$-$4$p~^{\rm{1}}$P$_{\rm{1}}$ & 5.082 & $-$30.638 & 0.06 & KUR07 & 7.58 & 7.57 & 0.13 & 0.12 & $-$0.01\\
16380.177 & 4$p~^{\rm{1}}$P$_{\rm{1}}$$-$3$d~^{\rm{1}}$P$^{\rm{o}}_{\rm{1}}$ & 5.863 & $-$30.495 & $-0.47$ & KUR07 & 7.03 & 7.03 & $-$0.95 & $-$0.95 & 0.00\\
16680.810 & 4$p~^{\rm{3}}$D$_{\rm{3}}$$-$3$d~^{\rm{3}}$D$^{\rm{o}}_{\rm{3}}$ & 5.984 & $-$30.357 & $-0.14$ & KUR07 & 7.48 & 7.45 & $-$0.17 & $-$0.20 & $-$0.03\\
16828.158 & 4$p~^{\rm{3}}$D$_{\rm{3}}$$-$3$d~^{\rm{3}}$D$^{\rm{o}}_{\rm{2}}$ & 5.984 & $-$30.357 & $-1.03$ & KUR07 & 7.41 & 7.41 & $-$1.13 & $-$1.13 & 0.00\\
mean      &                                                                  &       &           &         &       & 7.37 & 7.36 &         &         & \\
$\sigma$  &                                                                  &       &           &         &       & 0.24 & 0.24 &         &         &
\enddata
\
\tablecomments{References to the log\,$gf$ values are GAR73 : \citet{gar73}, KEL08 : \citet{kel08} and KUR07 : \citet{kur07}. The log\,$C_6$ values were calculated according to \citet{ans91,ans95} and \citet{bar00}. $\sigma$ refers to the statistical error. The log\,$gf'$ denotes that the $gf$-values were derived from the solar fits.}
\end{deluxetable*}

\subsubsection{Optical Atomic Line Data} \label{op_line}
We started with the same set of neutral Si optical lines used by \citet{shi09}. An examination shows that the line at 3905\,\AA\ is severely blended with a CH line, and the line at 4103\,\AA\ falls in the wing of H$_\delta$, while the line at 5690\,\AA\ is blended with an iron line. These three transitions were excluded from our Si abundance analysis. The adopted eight \ion{Si}{1} lines and line data are listed in Table\,\ref{tbl-3}. We also derived the solar Si abundance based on optical lines using the log\,$gf$ values from references. Although the mean Si abundance is consistent with the previous studies, the statistic error is also not satisfying, up to 0.1\,dex. To be consistent with the situation for infrared lines, we also present the $gf$-values determined from the solar spectrum fitting. The $C_6$ values were calculated according to \citet{ans91,ans95} and \citet{bar00}.

\subsection{NLTE Effects} \label{nlte_effects}
\subsubsection{Departures form LTE for the \ion{Si}{1} {\it{H}}-band lines} \label{nlte_h}

\begin{figure}
\includegraphics[scale=0.4,keepaspectratio=true,clip=true]{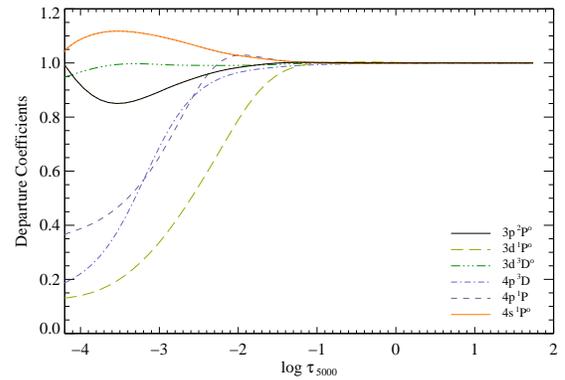}
\caption{Departure coefficients $b_{i} = N^{\rm{NLTE}}_{i}/N_{i}^{\rm{LTE}}$ as a function of the standard optical depth for HD 87. \label{fig1}}
\end{figure}

In Fig.\,\ref{fig1}, we present the departure coefficients ($b_{i}$) for the relevant \ion{Si}{1} levels for {\it{H}}-band transitions and \ion{Si}{2} ground state as a function of the optical depth at $\lambda$ = 5000\,\AA\ ($\tau_{5000}$) for the model atmosphere of \object{HD\,87}. Here, the departure coefficients ($b_{i}$) are defined as $b_{i} = n_{i}^{\rm{NLTE}}/n_{i}^{\rm{LTE}}$, where $n_{i}^{\rm{NLTE}}$ and $n_{i}^{\rm{LTE}}$ represent the statistical equilibrium (NLTE) and thermal (LTE) atomic level number densities, respectively. It is found that the departure coefficients for the level 3$d~^{3}{\rm D}^{o}$ of \ion{Si}{1} are near their thermal value ($b_{i} \sim 1$) and the level 4$s~^{\rm{1}}$P$^{\rm{o}}$ are overpopulated, while the other excitation levels, 4$p~^{\rm{1}}$P, 4$p~^{\rm{3}}$D, 3$d~^{1}{\rm P}^{\rm{o}}$, are underpopulated due to photon loss (see Fig.\,\ref{fig1} for details).

\begin{figure*}
\begin{center}
\includegraphics[scale=0.7,keepaspectratio=true,angle=90,clip=true]{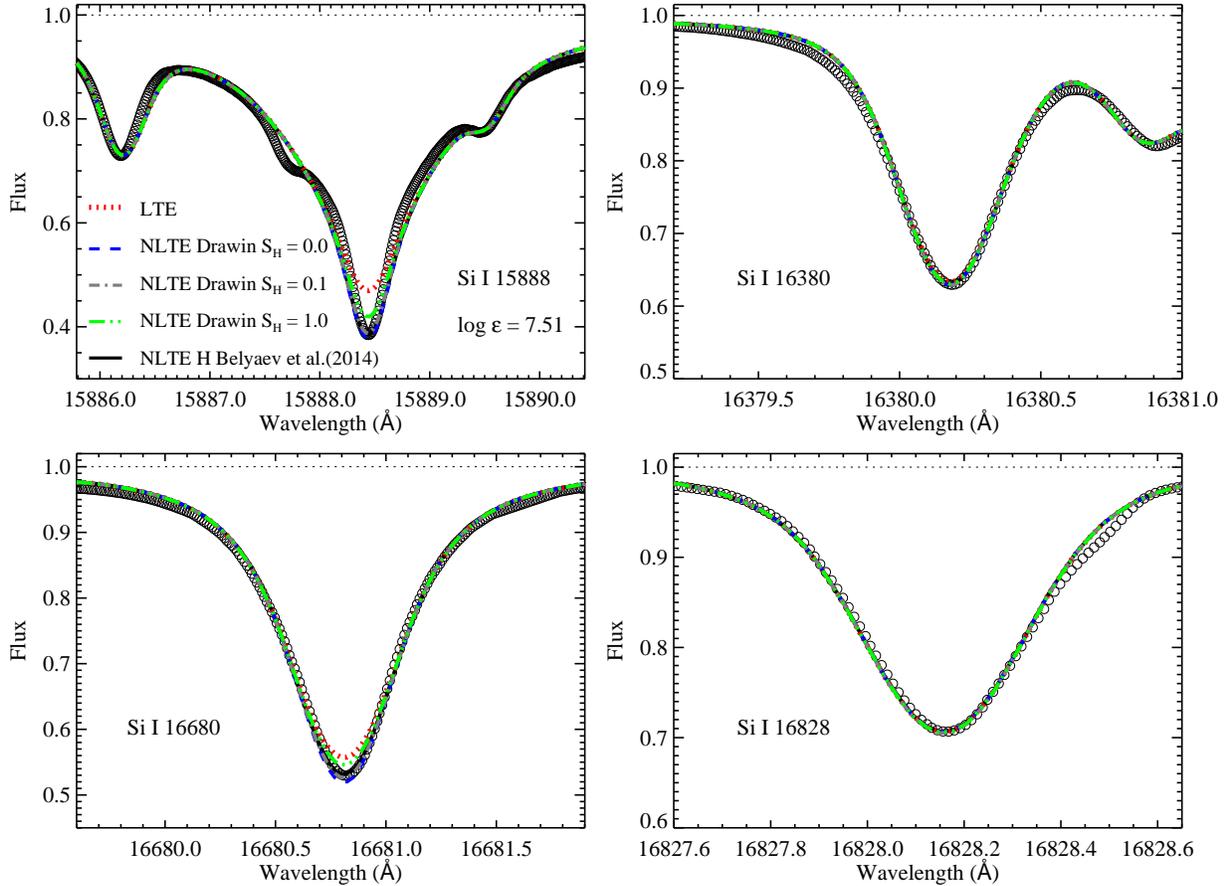}
\caption{{\it{H}}-band solar \ion{Si}{1} line profiles. The NLTE profiles with rates of collisions with hydrogen from \citet{bel14} and the Drawin recipe with S$_{\rm{H}} =$ 0.0, 0.1, 1.0, and LTE ones compared with the observed spectrum (open circles), where the NLTE profiles with the \citet{bel14} treatment refer to the best fits.}  \label{fig2}
\end{center}
\end{figure*}

\begin{figure*}
\begin{center}
\includegraphics[scale=0.7,keepaspectratio=true,angle=90,clip=true]{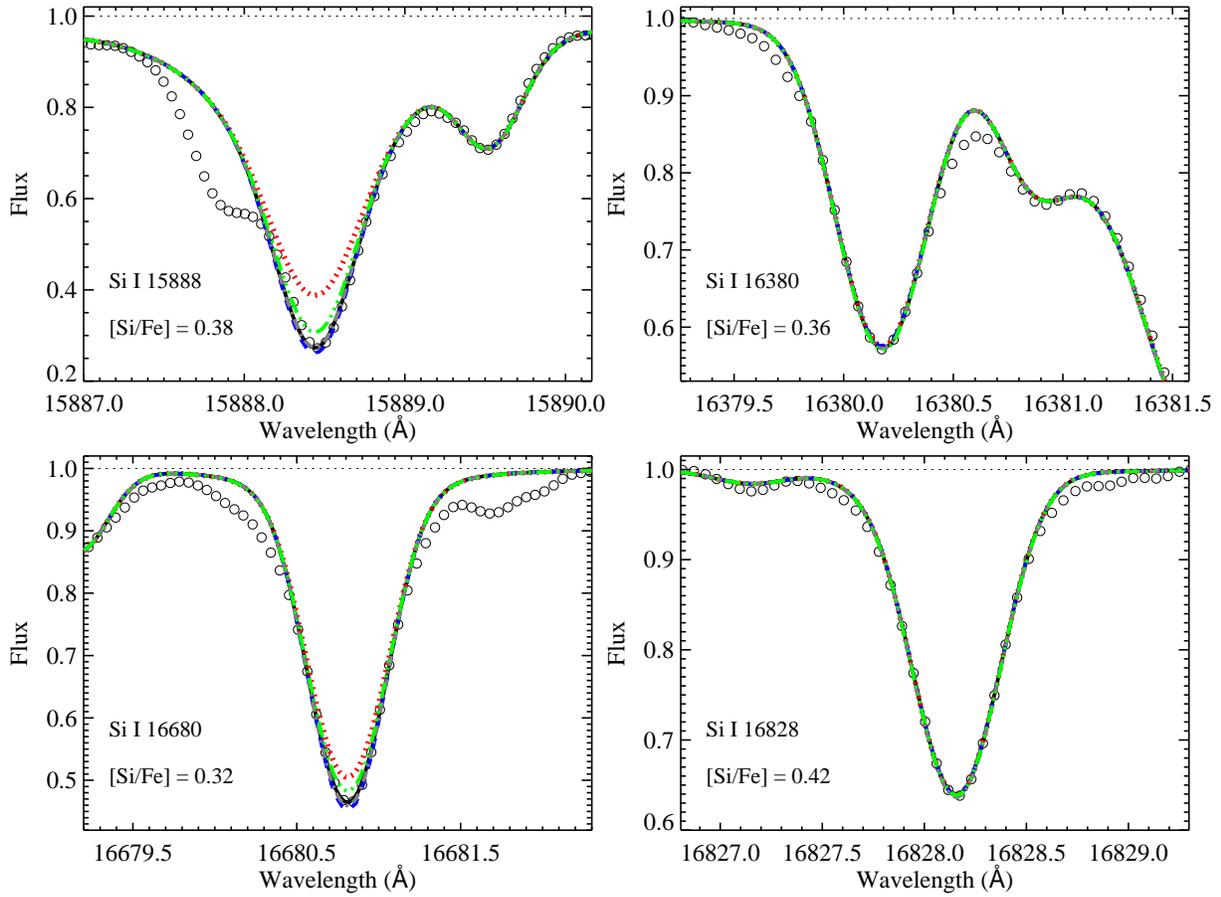}
\caption{Similar to Fig.\,\ref{fig2}, the NLTE and LTE profiles for Arcturus. Here the observed spectrum is from \citet{hin95}.}  \label{fig3}
\end{center}
\end{figure*}

\begin{figure}
\includegraphics[scale=0.45,keepaspectratio=true,clip=true]{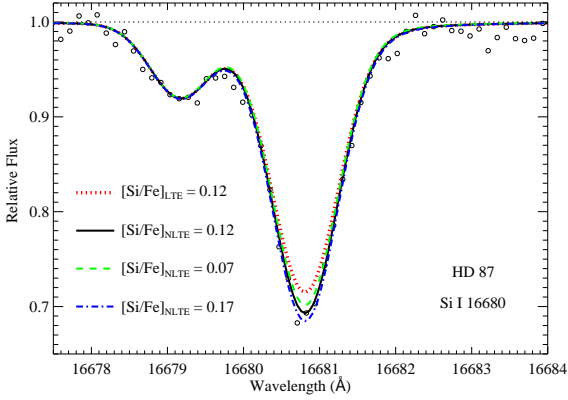}
\caption{Spectrum synthesis of the \ion{Si}{1} 16680\,\AA\ line for HD87. The open circles are the observed spectrum; the black solid line is the best-fitting NLTE line profile and the red dotted curve is the LTE profile with the same Si abundance; the other two lines are the synthetic spectra in NLTE with different [Si/Fe] (see the legend for details).  \label{fig4}}
\end{figure}

As the first test of our atomic model, we have analyzed optical lines for the Sun. We confirmed that the NLTE corrections for optical silicon lines are negligible \citep{shi08,wed01,ber13}. For the investigated four {\it{H}}-band Si lines, we found that the derived NLTE effects are also minor and the largest one is 0.03 dex (see Table\,\ref{tbl-3}). A comparison between the calculated {\it{H}}-band line profiles and the observed solar spectrum is shown in Fig.\,\ref{fig2}. In this figure, the NLTE (black solid) lines agree well with the observed spectrum for the strong lines at 15888 and 16680\,\AA, while the LTE (red dotted) line profiles are weaker. This issue is more obvious for \object{Arcturus}, as presented in Fig.\,\ref{fig3}. Fig.\,\ref{fig4} gives the synthetic profiles at 16680\,\AA\ under LTE and NLTE for \object{HD\,87}. The black solid line denotes the best fit to the observed spectrum in NLTE with a [Si/Fe] of 0.12\,dex. The red dotted curve is produced with the same [Si/Fe] in LTE, which is shallower in the line core.

In Table\,\ref{tbl-4}, we present the Si abundances derived from the individual {\it{H}}-band lines and the mean line-to-line scatter under NLTE and LTE for all sample stars. As indicated in this table, relative to LTE, NLTE obviously reduces the line-to-line scatter in the derived abundances for some stars. Taking \object{HD\,67447} as an example, the mean line-to-line scatter is reduced from 0.12 in LTE to 0.03\,dex in NLTE. Table\,\ref{tbl-5} gives the mean abundances along with the standard deviation. As shown in this table, the largest standard deviation in LTE is 0.12\,dex, however, it decreases to 0.07\,dex, when the NLTE effects are considered.

According to Table\,\ref{tbl-4}, the NLTE effects differ from line to line, and they are larger for strong lines. Among our four {\it{H}}-band lines, NLTE effects are relatively strong for the \ion{Si}{1} lines at 15888 and 16680\,\AA\ while they are smaller for the others. Table\,\ref{tbl-5} also shows the mean NLTE corrections for individual stars and the NLTE effects are from $-$0.1 to 0.0\,dex.

\begin{figure}[!htp]
\includegraphics[scale=0.45,keepaspectratio=true,clip=true]{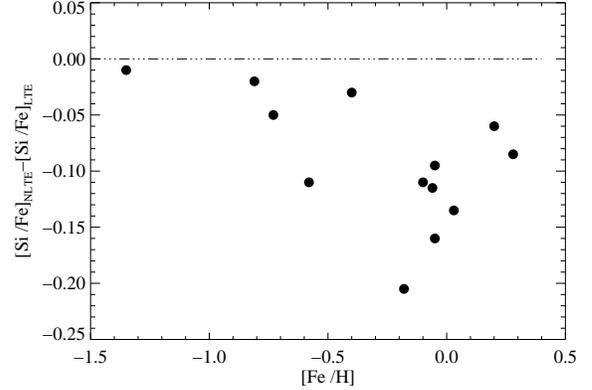}
\includegraphics[scale=0.45,keepaspectratio=true,clip=true]{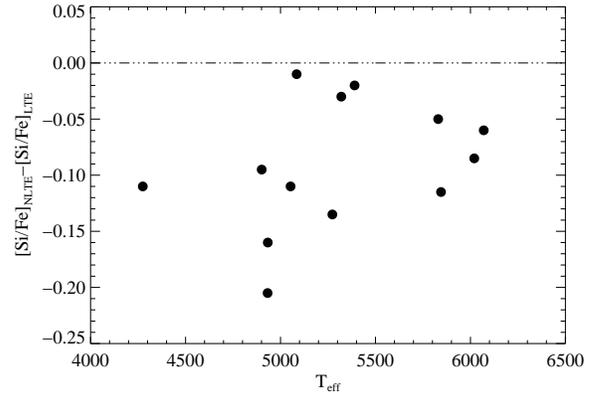}
\includegraphics[scale=0.45,keepaspectratio=true,clip=true]{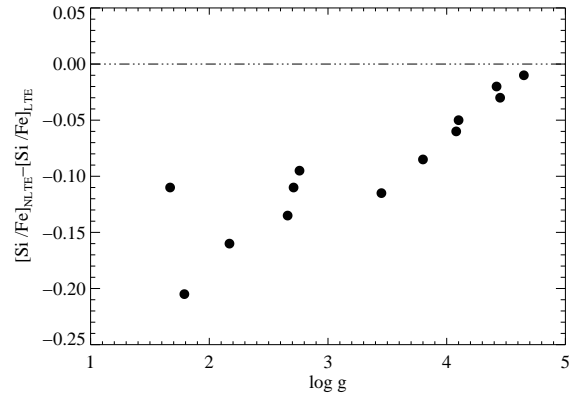}
\caption{The mean NLTE corrections for the two strong \ion{Si}{1} lines at 15888 and 16680\,\AA\ as functions of [Fe/H], $T_{\rm{eff}}$, and log\,$g$ respectively (from top to bottom). \label{fig5}}
\end{figure}

\begin{figure}[!htp]
\includegraphics[scale=0.45,keepaspectratio=true,clip=true]{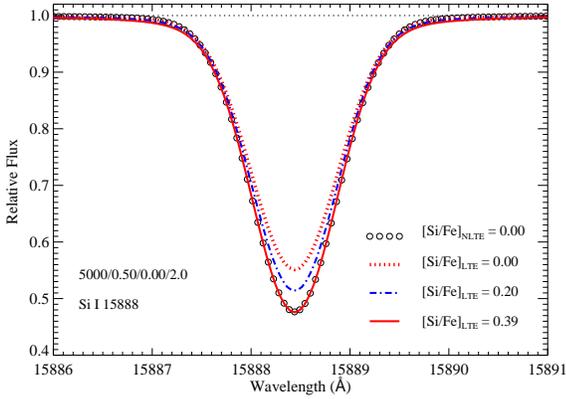}
\caption{The LTE and NLTE synthetic spectra of \ion{Si}{1} 15888\,\AA\ line with different [Si/Fe] values and the same parameters of $T_{\rm{eff}} =$ 5000\,K, log\,$g = 0.5$,  [Fe/H] $=$ 0.0, $\xi_t = 2.0$. [Si/Fe] $=$ 0.00, 0.20, 0.39\,dex for the LTE line profiles, while [Si/Fe] $=$ 0.00 for the NLTE calculation. \label{fig6}}
\end{figure}

To explore the dependency of the NLTE corrections on stellar parameters, we plotted the difference of the [Si/Fe] derived under NLTE and LTE assumptions for the strong silicon lines (15888 and 16680\,\AA) as functions of metallicity, effective temperature, and surface gravity in Fig.\,\ref{fig5}. It can be seen that the NLTE corrections of the {\it{H}}-band lines are negative, which means that the Si abundances would be overestimated under LTE; and that the NLTE effects are very sensitive to the surface gravity, the absolute corrections increase with the decreasing surface gravity, and the largest one reaches $\sim 0.2$\,dex for \object{HD\,58367}. Since surface gravity effects dominate, we do not see clear trends in the NLTE corrections with metallicity and effective temperature in these figures. 
In order to investigate the NLTE corrections for APOGEE data in extreme cases, we calculated the NLTE and LTE line profiles of the Si I line at 15888\,\AA\ with parameters $T_{\rm{eff}} =$ 5000\,K, [Fe/H] $=$ 0.0\,dex, log\,$g =$ 0.5, $\xi_t = $ 2.0\,km\,s$^{-1}$. As shown in Fig.\,\ref{fig6}, when [Si/Fe] under NLTE and LTE shares the same value, namely [Si/Fe] $=$ 0.0\,dex, the two profiles are different. By increasing [Si/Fe], the line cores of LTE spectra tend to be deeper and, until [Si/Fe] reaches 0.39\,dex, the LTE profile best fits the synthetic NLTE one. That is to say that in this extreme case, the NLTE correction can reach $\sim -$0.4\,dex.

\begin{figure*}
\plottwo{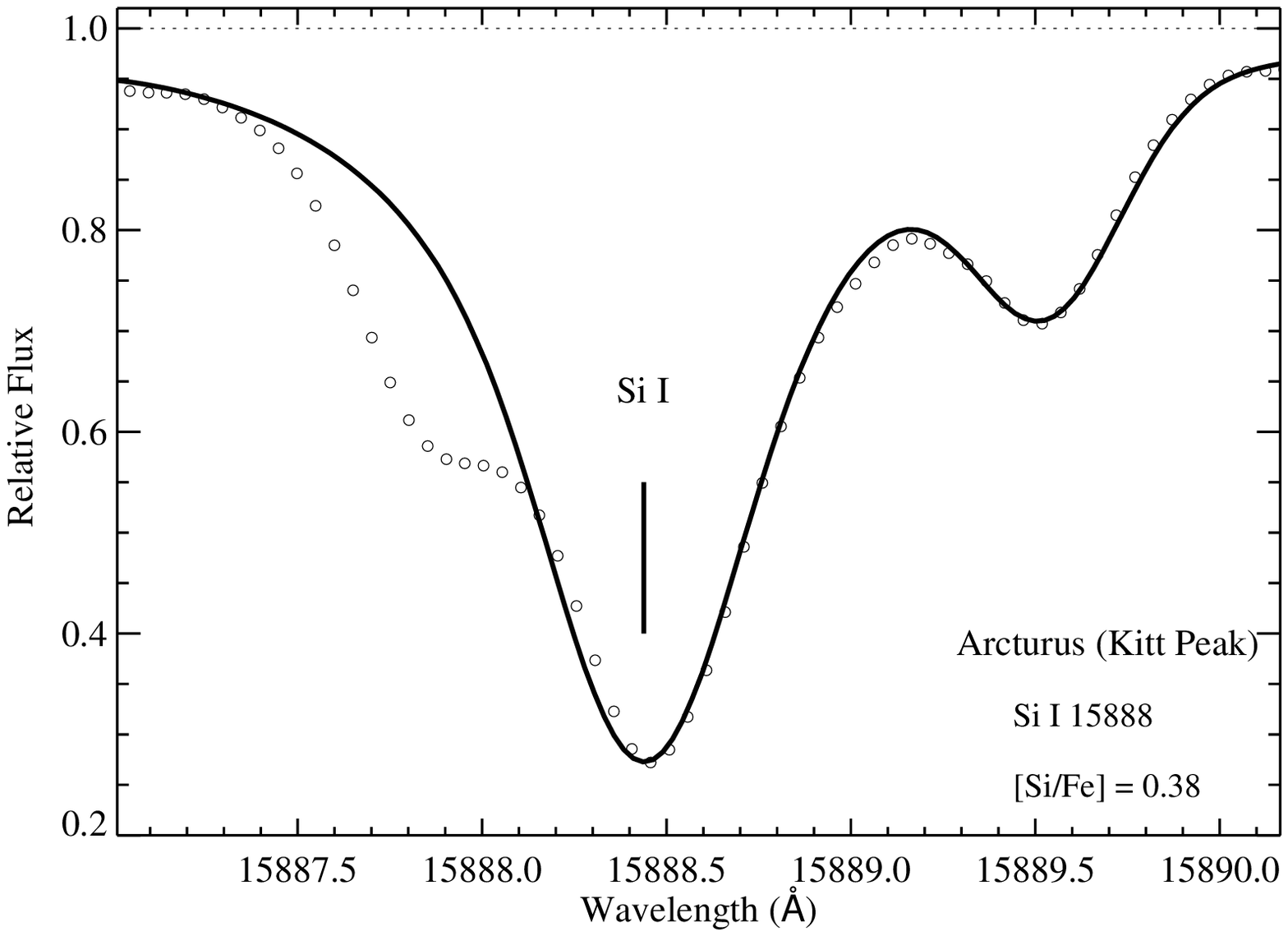}{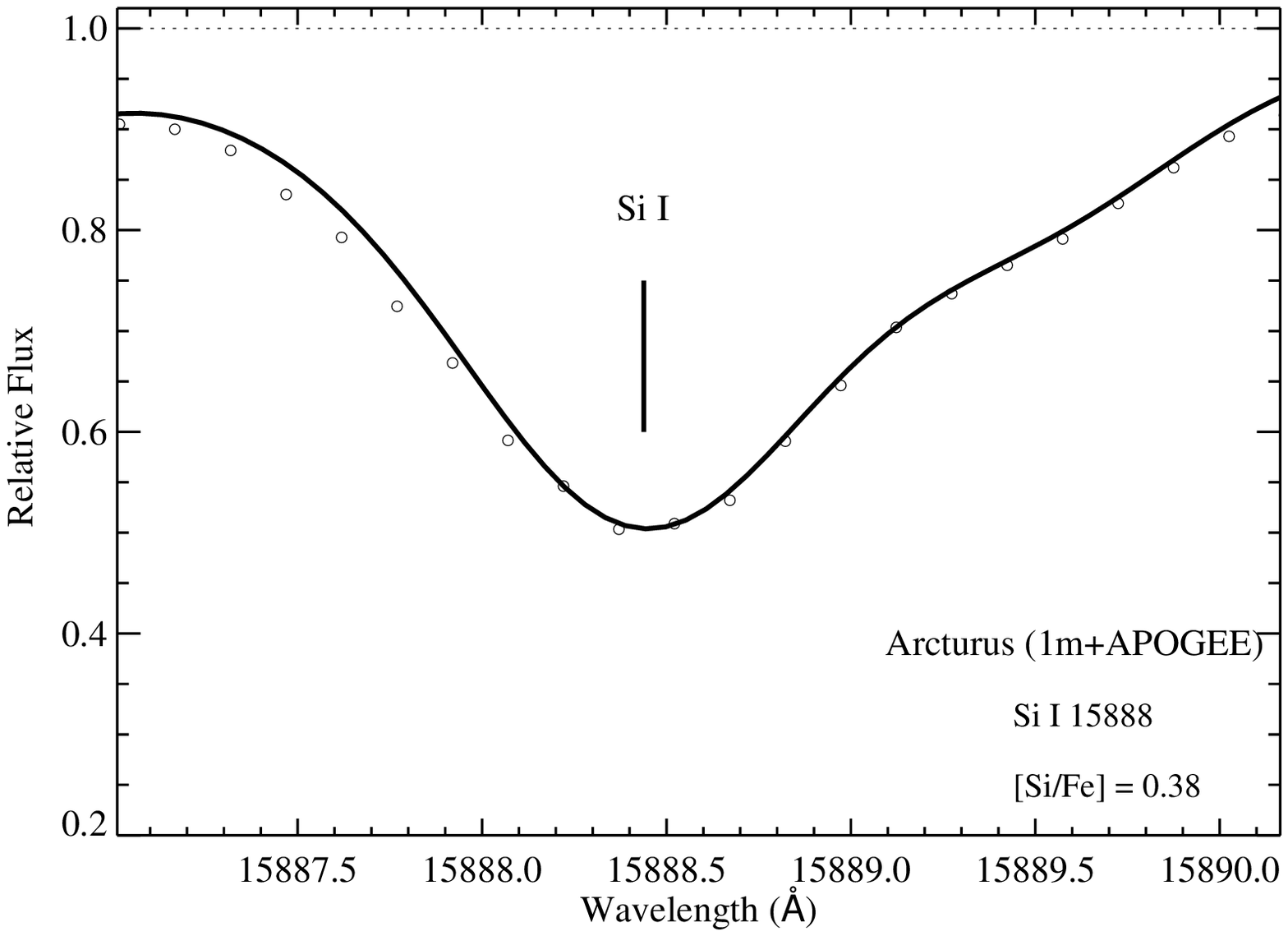}
\plottwo{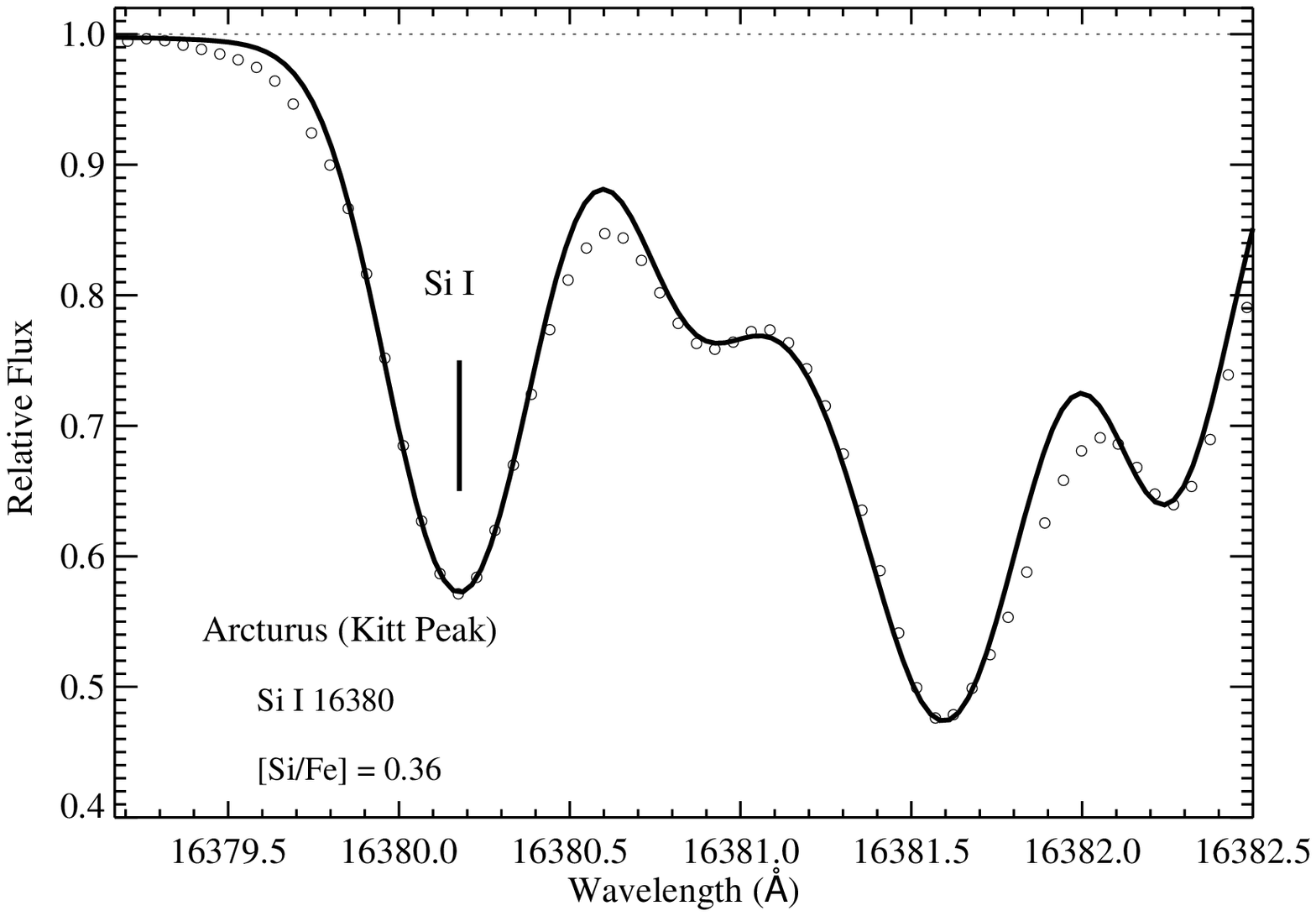}{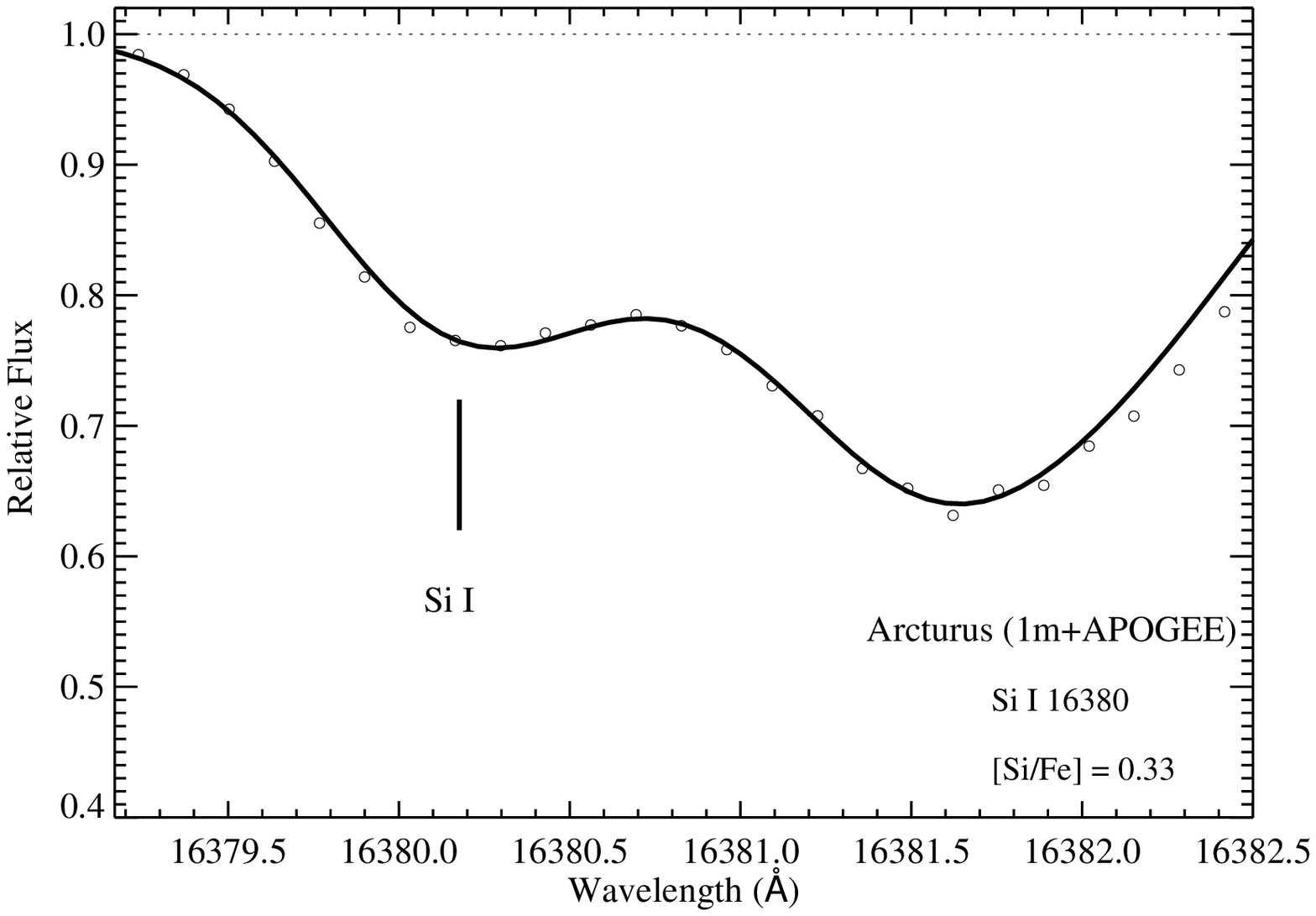}
\plottwo{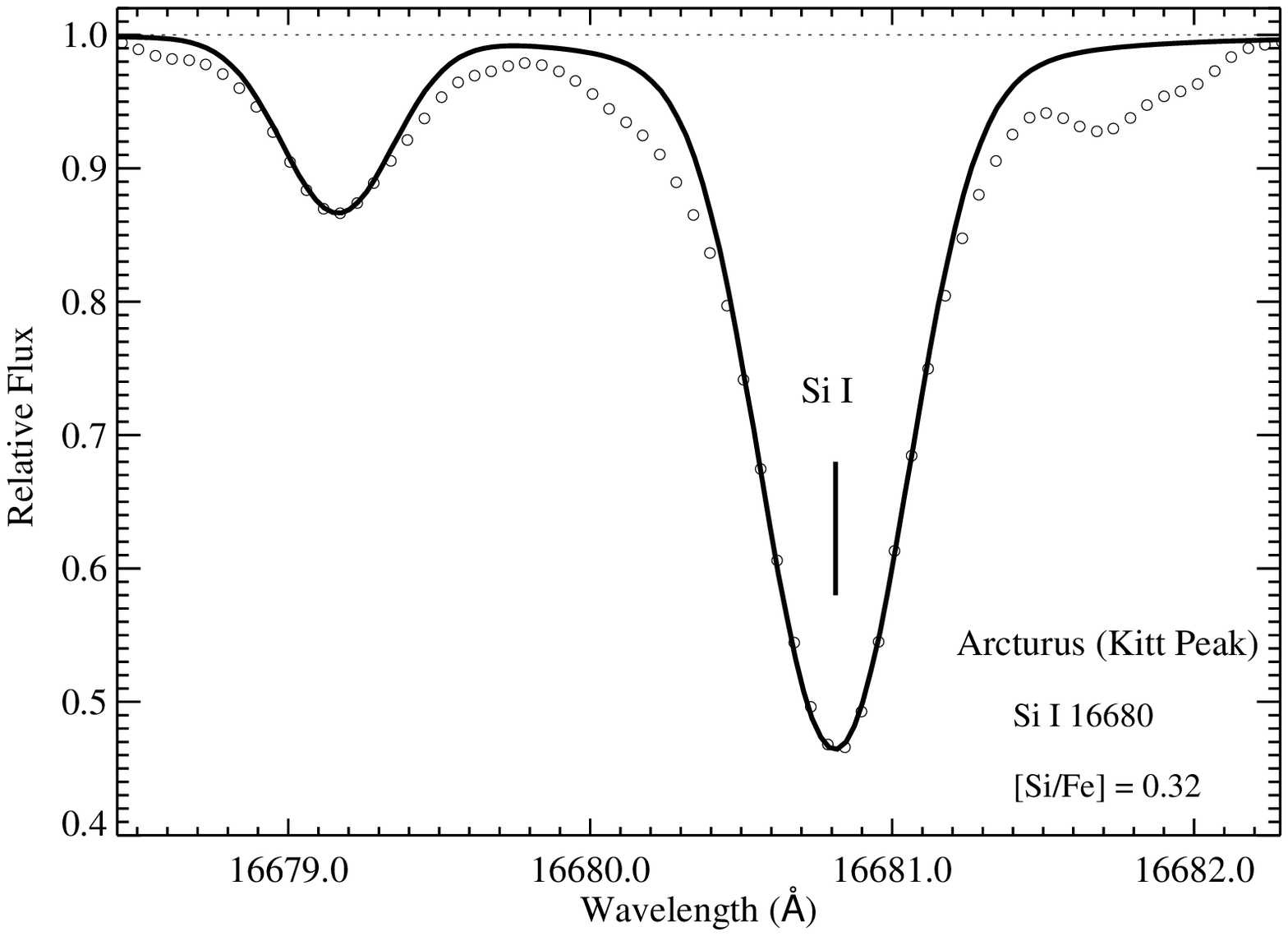}{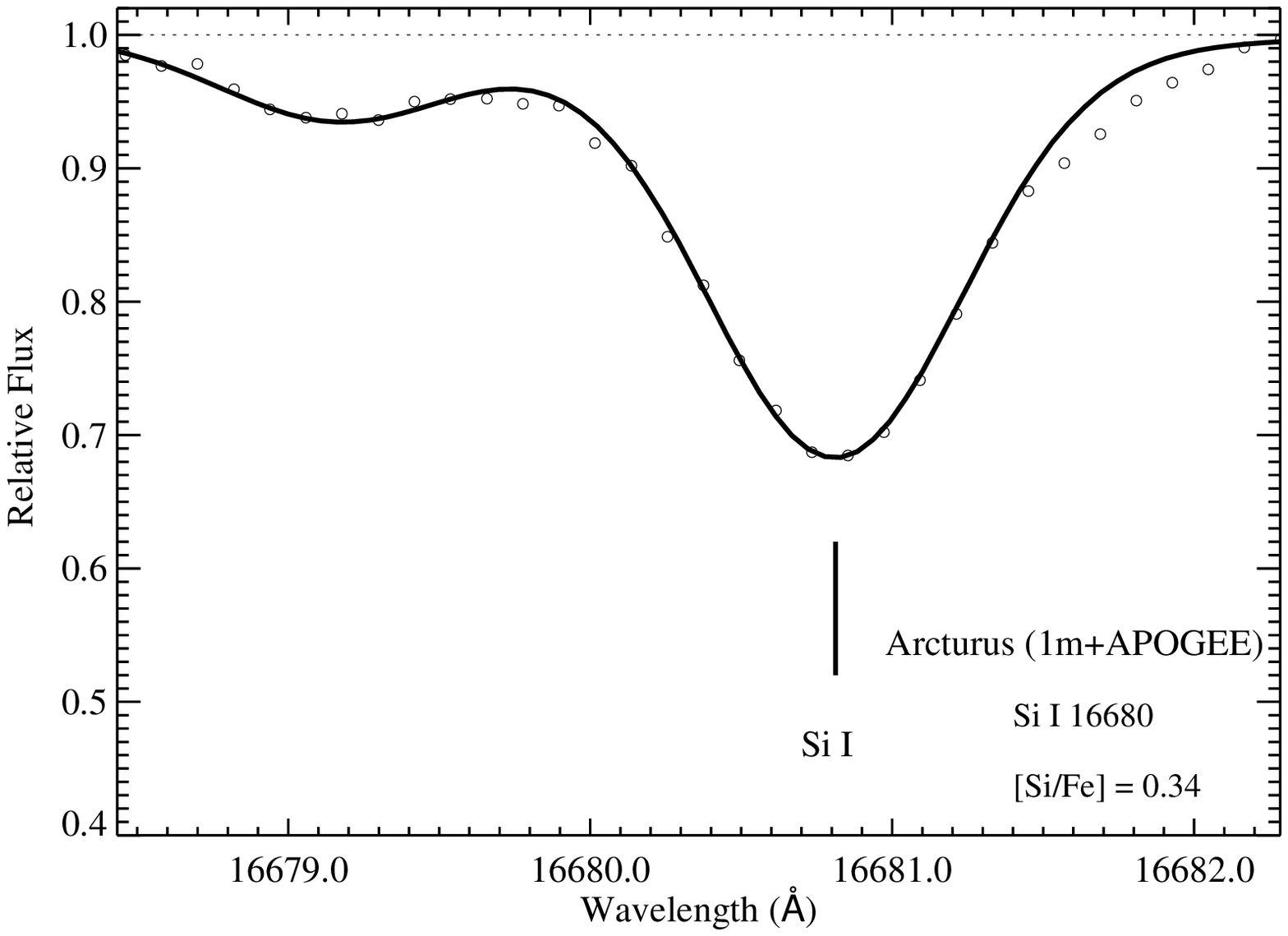}
\plottwo{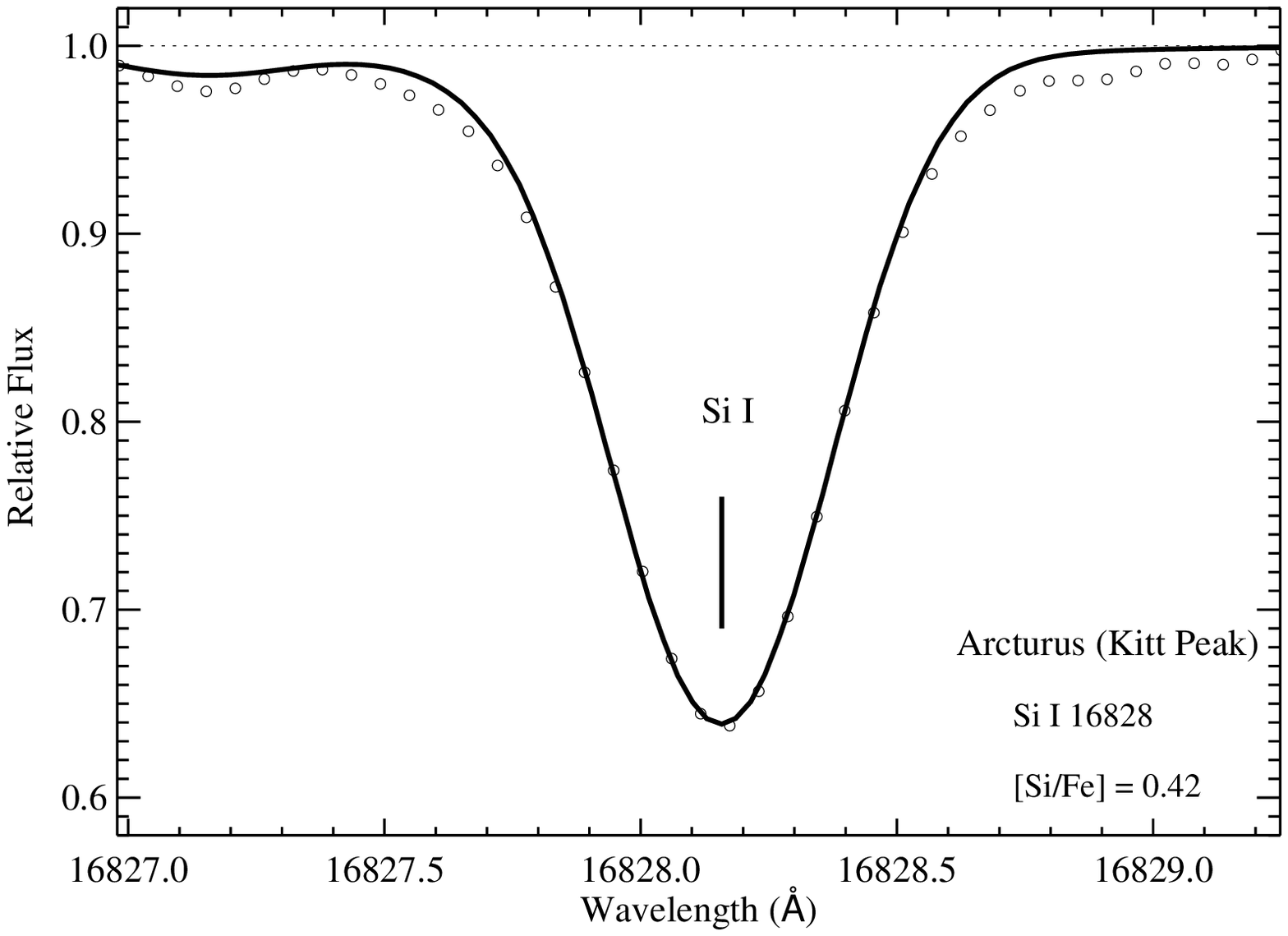}{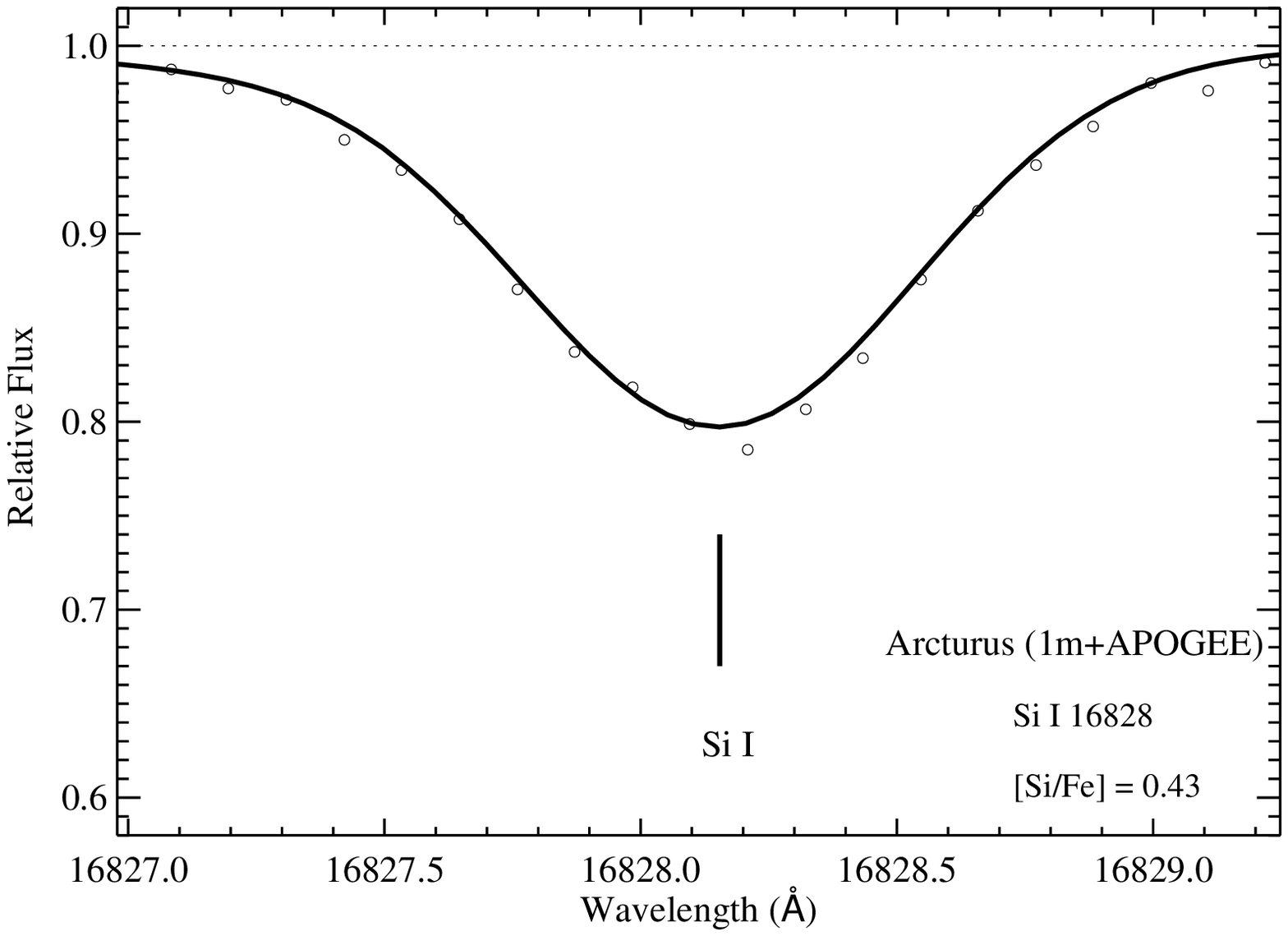}
\caption{The NLTE best fitting profiles (solid lines) of the four investigated \ion{Si}{1} lines in the Kitt Peak \citep{hin95} and 1m+APOGEE observed spectra of Arcturus (open circles). The left panel for the spectrum of Arcturus from \citet{hin95} while the right panel for the 1m+APOGEE spectrum.} \label{fig7}
\end{figure*}

To test whether consistent abundances are obtained from spectra acquired with different telescopes/instruments, we derived the Si abundance of \object{Arcturus} with the spectrum from \citet{hin95} (R $\sim$ 100,000) and the 1m$+$ APOGEE spectrum (R $\sim$ 22,500). Fig.\,\ref{fig7} demonstrates the best NLTE fitting profiles for the two observed spectra. The left panel is for the Arcturus spectrum from \citet{hin95} while the right one is for the 1m+APOGEE. Their results for individual lines are listed in Table\,\ref{tbl-4} and the mean ones in Table\,\ref{tbl-5}. The difference of abundances derived from individual lines between the two spectra is negligible, $\leq 0.02$\,dex. A consistent Si abundance is acquired for the same object from different telescopes/instruments.

\begin{deluxetable*}{lrrrrrrrrrrrrrr}
\tablecolumns{15}
\tabletypesize{\scriptsize}
\tablecaption{Stellar [Si/Fe] for the individual \ion{Si}{1} H-band lines under LTE  and NLTE analyses \label{tbl-4}}
\tablewidth{0pc}
\tablehead{
\colhead{} & \multicolumn{2}{c}{15888 (\AA)}  & \colhead{} & \multicolumn{2}{c}{16380 (\AA)}  & \colhead{} & \multicolumn{2}{c}{16680 (\AA)} & \colhead{} & \multicolumn{2}{c}{16828 (\AA)} & \colhead{} & \multicolumn{2}{c}{$\sigma_{\rm line}$}\\
\cline{2-3} \cline{5-6} \cline{8-9} \cline{11-12} \cline{14-15}\\
\colhead{Star} & \colhead{LTE} & \colhead{NLTE} & \colhead{}  & \colhead{LTE} & \colhead{NLTE} & \colhead{} & \colhead{LTE} & \colhead{NLTE} & \colhead{} & \colhead{LTE} & \colhead{NLTE} & \colhead{} & \colhead{LTE} & \colhead{NLTE}
}
\startdata
Arcturus\tablenotemark{a}  &   0.49  &   0.38  &  &    0.37 &  0.36   &  &   0.43  &  0.32   &  &  0.43   &  0.42  & & 0.06 & 0.05 \\
Arcturus\tablenotemark{b}  &   0.50  &   0.38  &  &    0.34 &  0.33   &  &   0.44  &  0.34   &  &  0.44   &  0.43  & & 0.08 & 0.06 \\
HD 87                      &   0.19  &  0.08   &  &    0.15 &  0.14   &  &   0.23  &  0.12   &  &  0.15   &  0.15  & & 0.05 & 0.04 \\
HD 6582                    &   0.22  &  0.19   &  &    0.26 &  0.25   &  &   0.26  &  0.25   &  &         &        & & 0.03 & 0.04 \\
HD 6920                    &   0.11  & $-$0.01 &  & $-$0.04 & $-$0.05 &  &   0.10  & $-$0.01 &  &         &        & & 0.10 & 0.03 \\
HD 22675                   &   0.12  &  0.02   &  &   0.07  &  0.06   &  &   0.15  &  0.06   &  &         &        & & 0.05 & 0.03 \\
HD 31501                   &   0.13  &  0.09   &  &   0.16  &  0.15   &  &   0.21  &  0.19   &  &         &        & & 0.05 & 0.07 \\
HD 58367                   &   0.26  &  0.03   &  &   0.04  &  0.04   &  &   0.31  &  0.13   &  &  0.12   & 0.13   & & 0.16 & 0.07 \\
HD 67447                   &   0.25  &  0.08   &  &   0.05  &  0.04   &  &   0.22  &  0.07   &  &  0.10   & 0.10   & & 0.12 & 0.03 \\
HD 102870                  & $-$0.02 & $-$0.09 &  & $-$0.07 & $-$0.08 &  & $-$0.04 & $-$0.09 &  & $-$0.08 & $-$0.08 & & 0.04 & 0.01 \\
HD 103095                  &   0.26  &  0.24   &  &   0.36  &  0.36   &  &   0.35  &   0.35  &  &         &        & & 0.07 & 0.08 \\
HD 121370                  &   0.11  &  0.02   &  &   0.15  &  0.14   &  &   0.22  &   0.14  &  &  0.14   &  0.14  & & 0.06 & 0.06 \\
HD 148816                  &   0.30  &  0.24   &  &   0.21  &  0.20   &  &   0.27  &   0.23  &  &         &        & & 0.06 & 0.03 \\
HD 177249                  &   0.20  &  0.07   &  &         &         &  &   0.20  &   0.06  &  &  0.08   &  0.08  & & 0.08 & 0.01
\enddata
\tablecomments{$\sigma_{\rm line}$ denotes the mean line-to-line scatter.}
\tablenotetext{a}{The H-band spectrum of Arcturus is from \citet{hin95}.}
\tablenotetext{b}{The H-band spectrum of Arcturus is the NMSU 1m $+$ APOGEE one.}
\end{deluxetable*}

\subsubsection{Departures form LTE for \ion{Si}{1} optical lines} \label{nlte_op}
We investigated the eight Si I optical lines described in Subsection \ref{op_line} for our sample stars. The mean Si abundances under LTE and NLTE are presented in Table\,\ref{tbl-5}. As shown in this table, the standard deviations are very small, less than 0.05\,dex for both LTE and NLTE abundances; the net NLTE correction for a given star is minor, with an absolute value less than 0.06 dex. Although the mean NLTE corrections are small, the NLTE effects are necessary for the strongest investigated \ion{Si}{1} lines, e.g. the largest NLTE correction for the line at 6155\,\AA\ reaches $\sim$ 0.1\,dex in our sample according to Table\,\ref{tbl-7}. The corrections could be greater in extreme cases.

\subsubsection{Comparison with the Optical Results and Discussions} \label{com}
\begin{figure}[!htp]
\includegraphics[scale=0.44,keepaspectratio=true,clip=true]{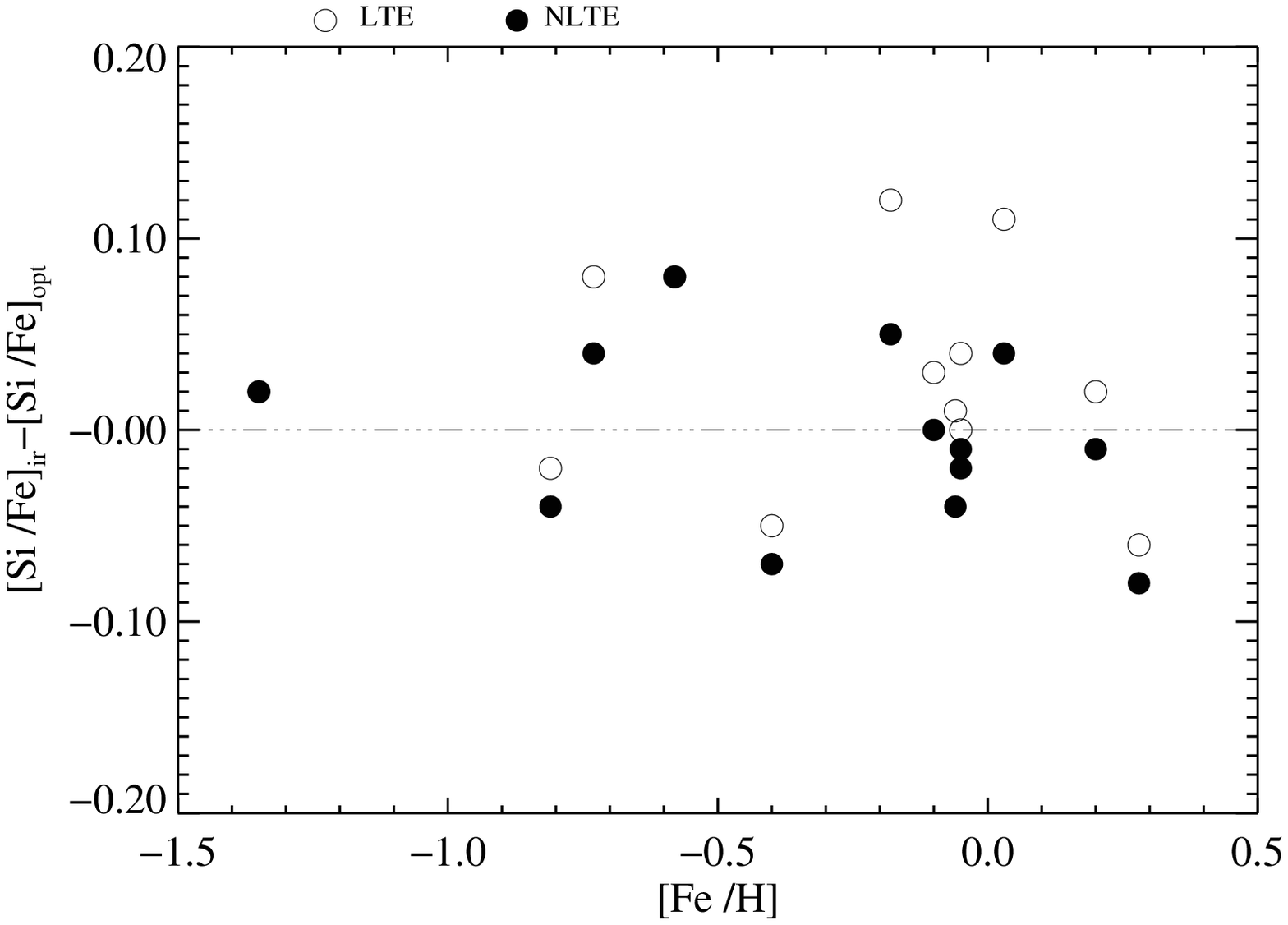}
\caption{The difference between the mean Si abundances derived from IR and optical lines under LTE (open circles) and NLTE (filled circles) assumptions. \label{fig8}}
\end{figure}

\begin{figure}[!htp]
\includegraphics[scale=0.44,keepaspectratio=true,clip=true]{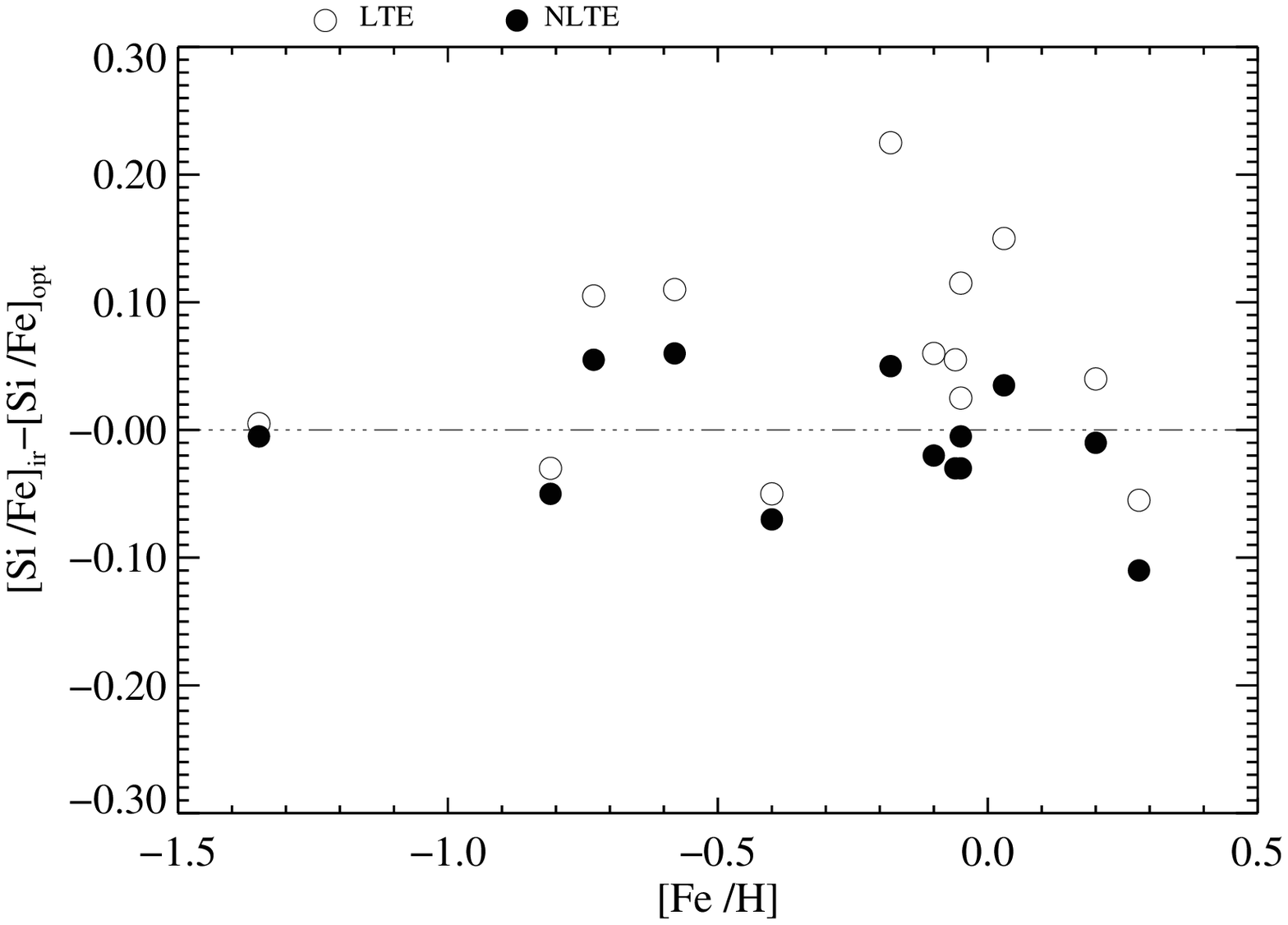}
\caption{The difference between the Si abundances derived from the two strong \ion{Si}{1} lines at 15888 and 16680\,\AA\ and the optical lines under LTE (open circles) and NLTE (filled circles). \label{fig9}}
\end{figure}

For our sample stars, the differences between the mean Si abundances derived from IR and from optical spectra are shown against the metallicity in Fig.\,\ref{fig8}. In this figure, open circles denote the differences in LTE while filled circles indicate the NLTE results. As one can see, the differences between LTE and NLTE are small (less than 0.1\,dex) and the derived Si abundances from the {\it{H}}-band spectral lines agree better with those from optical lines in NLTE than in LTE. Since the NLTE effects are larger for strong lines, it is interesting to see whether the Si abundances derived when only {\it{H}}-band strong lines at 15888 and 16680\,\AA\ available are still consistent with the ones from optical lines. The differences between the abundances derived from the two strong {\it{H}}-band lines and from optical lines are depicted in Fig.\,\ref{fig9}. Similar to Fig.\,\ref{fig8}, the NLTE Si abundances from the strong {\it{H}}-band lines are consistent with those from optical lines, while the differences become as large as 0.2\,dex in LTE.

\begin{deluxetable*}{lrcrrrrrrrr}
\tabletypesize{\scriptsize}
\tablecaption{Stellar silicon LTE and NLTE abundances\label{tbl-5}}
\tablewidth{0pt}
\tablehead{
\colhead{Star} & \colhead{$T_{\rm{eff}}$} & \colhead{log\,$g$} & \colhead{[Fe/H]} & \colhead{$\xi_t$} & \colhead{[{Si I}$_{LTE}$/Fe](ir)} & \colhead{[{Si I}$_{NLTE}$/Fe](ir)} & \colhead{$\rm\Delta_{ir}$} & \colhead{[Si I$_{LTE}$/Fe](opt)} & \colhead{[{Si I}$_{NLTE}$/Fe](opt)} & \colhead{$\rm\Delta_{opt}$}
}
\startdata
Arcturus\tablenotemark{a} & 4275 & 1.67 & $-$0.58 & 1.60 & 0.43$\pm$0.05  & 0.37$\pm$0.04 & $-$0.06 & 0.35$\pm$0.03 & 0.29$\pm$0.02 & $-$0.06\\
Arcturus\tablenotemark{b} & 4275 & 1.67 & $-$0.58 & 1.60 & 0.43$\pm$0.07  & 0.37$\pm$0.05 & $-$0.06\\
HD 87     & 5053 & 2.71 & $-$0.10 & 1.35 & 0.18$\pm$0.04 & 0.12$\pm$0.03 & $-$0.06 &0.15$\pm$0.03 & 0.12$\pm$0.01 & $-$0.03\\
HD 6582   & 5390 & 4.42 & $-$0.81 & 0.90 & 0.25$\pm$0.02 & 0.23$\pm$0.03 & $-$0.02 & 0.27$\pm$0.02 & 0.27$\pm$0.02 & 0.00\\
HD 6920   & 5845 & 3.45 & $-$0.06 & 1.40 & 0.06$\pm$0.08 & $-0.02$$\pm$0.02 & $-$0.08 & 0.05$\pm$0.05 & 0.02$\pm$0.03 & $-$0.03\\
HD 22675  & 4901 & 2.76 & $-$0.05 & 1.30 & 0.11$\pm$0.04 & 0.05$\pm$0.02 & $-$0.06 & 0.11$\pm$0.04 & 0.07$\pm$0.02 & $-$0.04\\
HD 31501  & 5320 & 4.45 & $-$0.40 & 1.00 & 0.17$\pm$0.04 & 0.14$\pm$0.05 & $-$0.03 & 0.22$\pm$0.02 & 0.21$\pm$0.02 & $-$0.01\\
HD 58367  & 4932 & 1.79 & $-$0.18 & 2.00 & 0.18$\pm$0.12 & 0.08$\pm$0.06 & $-$0.10 & 0.16$\pm$0.05 & 0.13$\pm$0.02 & $-$0.03\\
HD 67447  & 4933 & 2.17 & $-$0.05 & 2.12 & 0.16$\pm$0.10 & 0.07$\pm$0.02 & $-$0.09 & 0.12$\pm$0.04 & 0.08$\pm$0.02 & $-$0.04\\
HD 102870 & 6070 & 4.08 &    0.20 & 1.20 & $-$0.05$\pm$0.03 & $-$0.09$\pm$0.01 & $-$0.04 & $-$0.07$\pm$0.02 & $-$0.08$\pm$0.02 & $-$0.01\\
HD 103095 & 5085 & 4.65 & $-$1.35 & 0.80 & 0.32$\pm$0.06 & 0.32$\pm$0.07 & 0.00 & 0.30$\pm$0.04 & 0.30$\pm$0.04 & 0.00\\
HD 121370 & 6020 & 3.80 &    0.28 & 1.40 & 0.16$\pm$0.05 & 0.11$\pm$0.06 & $-$0.05 & 0.22$\pm$0.05 & 0.19$\pm$0.03 & $-$0.03\\
HD 148816 & 5830 & 4.10 & $-$0.73 & 1.40 & 0.26$\pm$0.05 & 0.22$\pm$0.02 & $-$0.04 & 0.18$\pm$0.03 & 0.18$\pm$0.03 & 0.00\\
HD 177249 & 5273 & 2.66 &    0.03 & 1.65 & 0.16$\pm$0.07 & 0.07$\pm$0.01 & $-$0.09 & 0.05$\pm$0.04 & 0.03$\pm$0.02 & $-$0.02
\enddata
\tablecomments{$\rm\Delta_{ir}$ and $\rm\Delta_{opt}$ stand for the NLTE effects ($\Delta$ $=$ $\rm{log}\,\varepsilon_{NLTE}$ $-$ $\rm{log}\,\varepsilon_{LTE}$) derived from IR and optical spectra respectively.}
\tablenotetext{a}{The {\it{H}}-band spectrum of Arcturus is from \citet{hin95}.}
\tablenotetext{b}{The {\it{H}}-band spectrum of Arcturus is the NMSU 1m $+$ APOGEE one.}
\end{deluxetable*}

\section{CONCLUSIONS} \label{con}
The main purpose of this work is to test the validity of the Si atomic model for {\it{H}}-band line formation, and to investigate the NLTE effects on Si spectral lines based on high S/N IR {\it{H}}-band spectra. A sample of 13 FGK dwarfs and giants was selected, and the Si abundances were derived from both {\it{H}}-band and optical lines under LTE and NLTE.

After careful analysis, we conclude that:
\begin{itemize}
\item With a NLTE analysis, the absolute differences between the mean Si abundances from the {\it{H}}-band and from optical lines are less than 0.1\,dex for the sample stars, which suggests that our Si atomic model can be applied for investigating the formation of the {\it{H}}-band \ion{Si}{1} lines.
\item The NLTE effects differ from line to line. The strong \ion{Si}{1} lines at 15888 and 16680\,\AA\ need large NLTE corrections, while the other two lines show weaker NLTE effects. Thus, it is not surprising that the NLTE silicon abundance shows a smaller line-to-line scatter than the LTE one for some stars in this analysis. The NLTE corrections reach $\sim -$0.2\,dex for the strongest \ion{Si}{1} line among our sample. It can be up to $\sim -$0.4\,dex for the extreme cases of APOGEE targets (log\,$g$ $\sim$ 0.5). Thus it should be considered in the abundance analysis, especially for the cases where only strong lines are available.

\item The NLTE effects are sensitive to the surface gravity, and increase with decreasing surface gravity.

\item The NLTE corrections for the investigated {\it{H}}-band lines are negative, which means that the Si abundances derived with LTE assumption will be overestimated.
\end{itemize}

To the best of our knowledge, this work is the first NLTE investigation of the {\it{H}}-band Si spectral lines. The NLTE corrections of strong lines range from $-0.2$ to $-$0.1\,dex for giant stars in our sample. In extreme cases of APOGEE targets, the correction could be up to $-$0.4\,dex.  Thus they may have a significant impact on the Si abundances derived from APOGEE observations. Motivated by these results, the APOGEE team is planning to pursue more extended NLTE calculations in the coming years.

\acknowledgments

This research is supported by National Key Basic Research Program of China 2014CB845700, and by the National Natural Science Foundation of China under grant Nos. 11321064, 11233004, 11390371, 11473033, 11428308, U1331122. CAP is thankful to the Spanish MINECO for support through grant AYA2014-56359-P.

We acknowledge the support of the staff of the Xinglong 2.16m telescope. This work was partially Supported by the Open Project Program of the Key Laboratory of Optical Astronomy, National Astronomical Observatories, Chinese Academy of Sciences. We thank Yoichi Takeda, Bun'ei Sato and Yujuan Liu for providing us the optical data. The authors thank the anonymous referee for comments that helped to improve the manuscript.

This research uses services or data provided by the NOAO Science Archive. NOAO is operated by the Association of Universities for Research in Astronomy (AURA), Inc. under a cooperative agreement with the National Science Foundation.

This work has made use of the VALD database, operated at Uppsala University, the Institute of Astronomy RAS in Moscow, and the University of Vienna.

Funding for SDSS-III has been provided by the Alfred P. Sloan Foundation, the Participating Institutions, the National Science Foundation, and the U.S. Department of Energy Office of Science. The SDSS-III web site is \texttt{http://www.sdss3.org/}.

SDSS-III is managed by the Astrophysical Research Consortium for the Participating Institutions of the SDSS-III Collaboration including the University of Arizona, the Brazilian Participation Group, Brookhaven National Laboratory, Carnegie Mellon University, University of Florida, the French Participation Group, the German Participation Group, Harvard University, the Instituto de Astrofisica de Canarias, the Michigan State/Notre Dame/JINA Participation Group, Johns Hopkins University, Lawrence Berkeley National Laboratory, Max Planck Institute for Astrophysics, Max Planck Institute for Extraterrestrial Physics, New Mexico State University, New York University, Ohio State University, Pennsylvania State University, University of Portsmouth, Princeton University, the Spanish Participation Group, University of Tokyo, University of Utah, Vanderbilt University, University of Virginia, University of Washington, and Yale University.

\clearpage
\appendix
\clearpage

\begin{deluxetable}{lccrlccrr}
\tablecolumns{4}
\tabletypesize{\scriptsize}
\tablecaption{Line data, iron abundances derived from the solar spectrum and equivalent widths of the solar lines  \label{tbl-6}}
\tablewidth{0pt}
\tablehead{
\colhead{$\lambda$} & \colhead{$\chi$} & \colhead{log\,$C_6$} & \colhead{log\,$gf$}  & \colhead{Ref.}  & \colhead{$\rm{log}\,\varepsilon_{\sun}Fe$} & \colhead{$\rm{log}\,\varepsilon_{\sun}Fe$} & \colhead{log\,$gf'$} & \colhead{EW}\\
\colhead{} & \colhead{} & \colhead{} & \colhead{}  & \colhead{}  & \colhead{LTE} & \colhead{NLTE} & \colhead{} & \colhead{}\\
\colhead{(\AA)} & \colhead{(eV)} & \colhead{} & \colhead{}  & \colhead{}  & \colhead{(dex)} & \colhead{(dex)} & \colhead{} & \colhead{(m\AA)}
}
\startdata
\ion{Fe}{1}\\																														
4661.534 	&	4.558 	&	$-$29.481 	&	$-$1.27 	&	FUH88	&	7.57 	&	7.61 	&	$-$1.16 	&	40.5 	\\
4808.149 	&	3.251 	&	$-$31.464 	&	$-$2.79 	&	FUH88	&	7.66 	&	7.70 	&	$-$2.59 	&	29.5 	\\
4885.430 	&	3.882 	&	$-$30.173 	&	$-$1.02 	&	KUR14	&	7.49 	&	7.55 	&	$-$0.97 	&	91.3 	\\
5223.186 	&	3.635 	&	$-$31.165 	&	$-$1.78 	&	BRI91	&	7.05 	&	7.09 	&	$-$2.19 	&	31.0 	\\
5242.497 	&	3.634 	&	$-$31.248 	&	$-$0.97 	&	BRI91	&	7.56 	&	7.52 	&	$-$0.95 	&	90.3 	\\
5379.579 	&	4.154 	&	$-$31.242 	&	$-$1.51 	&	BRI91	&	7.57 	&	7.57 	&	$-$1.44 	&	63.5 	\\
5398.279 	&	4.371 	&	$-$30.155 	&	$-$0.67 	&	FUH88	&	7.55 	&	7.59 	&	$-$0.58 	&	78.8 	\\
5522.449 	&	4.217 	&	$-$30.457 	&	$-$1.55 	&	FUH88	&	7.63 	&	7.68 	&	$-$1.37 	&	44.9 	\\
5546.506 	&	4.434 	&	$-$30.356 	&	$-$1.31 	&	FUH88	&	7.68 	&	7.74 	&	$-$1.07 	&	52.7 	\\
5618.633 	&	4.386 	&	$-$30.475 	&	$-$1.28 	&	BRI91	&	7.49 	&	7.55 	&	$-$1.23 	&	52.2 	\\
5651.469 	&	4.386 	&	$-$30.264 	&	$-$2.00 	&	FUH88	&	7.77 	&	7.78 	&	$-$1.72 	&	19.5 	\\
5679.023 	&	4.186 	&	$-$30.040 	&	$-$0.92 	&	FUH88	&	7.72 	&	7.78 	&	$-$0.64 	&	65.0 	\\
5793.915 	&	4.220 	&	$-$30.505 	&	$-$1.70 	&	FUH88	&	7.58 	&	7.63 	&	$-$1.57 	&	35.5 	\\
5853.148 	&	1.485 	&	$-$31.586 	&	$-$5.28 	&	FUH88	&	7.64 	&	7.67 	&	$-$5.11 	&	8.1 	\\
5855.077 	&	4.608 	&	$-$30.189 	&	$-$1.48 	&	BAR94	&	7.43 	&	7.48 	&	$-$1.50 	&	23.3 	\\
5929.677 	&	4.548 	&	$-$30.305 	&	$-$1.41 	&	FUH88	&	7.71 	&	7.77 	&	$-$1.14 	&	41.7 	\\
6024.058 	&	4.548 	&	$-$30.358 	&	$-$0.12 	&	FUH88	&	7.66 	&	7.70 	&	   0.08 	&	127.5 	\\
6078.491 	&	4.796 	&	$-$29.749 	&	$-$0.32 	&	KUR14	&	7.47 	&	7.52 	&	$-$0.30 	&	84.6 	\\
6079.009 	&	4.652 	&	$-$30.237 	&	$-$1.12 	&	FUH88	&	7.64 	&	7.70 	&	$-$0.92 	&	48.8	\\
6151.623 	&	2.176 	&	$-$31.538 	&	$-$3.30 	&	FUH88	&	7.53 	&	7.55 	&	$-$3.25 	&	51.6 	\\
6173.335 	&	2.223 	&	$-$31.523 	&	$-$2.88 	&	FUH88	&	7.56 	&	7.58 	&	$-$2.80 	&	70.1 	\\
6200.321 	&	2.608 	&	$-$31.279 	&	$-$2.44 	&	FUH88	&	7.59 	&	7.59 	&	$-$2.35 	&	75.2 	\\
6240.646 	&	2.223 	&	$-$31.450 	&	$-$3.23 	&	BAR91	&	7.44 	&	7.46 	&	$-$3.27 	&	48.7 	\\
6322.686 	&	2.588 	&	$-$31.296 	&	$-$2.43 	&	FUH88	&	7.60 	&	7.60 	&	$-$2.33 	&	77.6 	\\
6335.331 	&	2.198 	&	$-$31.546 	&	$-$2.18 	&	BRI91	&	7.46 	&	7.46 	&	$-$2.22 	&	103.3 	\\
6481.877 	&	2.279 	&	$-$31.420 	&	$-$2.98 	&	FUH88	&	7.58 	&	7.60 	&	$-$2.88 	&	65.7 	\\
6593.871 	&	2.433 	&	$-$31.375 	&	$-$2.42 	&	FUH88	&	7.62 	&	7.63 	&	$-$2.29 	&	98.7 	\\
6726.666 	&	4.607 	&	$-$30.256 	&	$-$1.09 	&	KUR14	&	7.56 	&	7.63 	&	$-$0.96 	&	50.2 	\\
6839.831 	&	2.559 	&	$-$31.346 	&	$-$3.45 	&	FUH88	&	7.55 	&	7.58 	&	$-$3.37 	&	30.3 	\\
6857.250 	&	4.076 	&	$-$30.895 	&	$-$2.15 	&	FUH88	&	7.56 	&	7.61 	&	$-$2.04 	&	23.4 	\\
mean	    &		    &		        &		    &		    &	7.56 	&	7.60 	&		    &		    \\
$\sigma$    &		    &		        &		    &		    &	0.13 	&	0.13 	&		    &		    \\
\hline\\
\ion{Fe}{2}\\
4508.288 	&	2.856 	&	$-$31.971	&	$-$2.25 	&	RYA99	&	7.48 	&	7.48 	&	$-$2.27 	&	77.6 	\\
5264.808 	&	3.230 	&	$-$31.977	&	$-$3.12 	&	RYA99	&	7.53 	&	7.53 	&	$-$3.09 	&	103.3 	\\
5414.073 	&	3.221 	&	$-$31.976	&	$-$3.54 	&	RYA99	&	7.45 	&	7.45 	&	$-$3.60 	&	65.7 	\\
5991.376 	&	3.153 	&	$-$31.983	&	$-$3.54 	&	BLA80	&	7.43 	&	7.43 	&	$-$3.61 	&	98.7 	\\
6149.258 	&	3.889 	&	$-$32.048	&	$-$2.72 	&	BLA80	&	7.49 	&	7.49 	&	$-$2.73 	&	50.2 	\\
6456.383 	&	3.903 	&	$-$31.979	&	$-$2.10 	&	BLA80	&	7.54 	&	7.54 	&	$-$2.07 	&	30.3 	\\
mean	    &		    &		    &		    &		    &	7.49 	&	7.49 	&		    &        \\
$\sigma$	&		    &		    &		    &		    &	0.04 	&	0.04 	&		    &
\enddata
\tablecomments{References to the log\,$gf$ values are FUH88: \citet{fuh88}, KUR14: \citet{kur14}, BRI91: \citet{bri91}, BAR94: \citet{bar94}, BAR91: \citet{bar91}, RYA99: \citet{rya99} and BLA80: \citet{bla80}. The log\,$C_6$ values were calculated according to \citet{ans91,ans95} and \citet{bar00}. $\sigma$ refers to the statistical error. The log\,$gf'$ denotes that the $gf$-values were derived from the NLTE solar fits.}
\end{deluxetable}

\clearpage
\begin{landscape}
\begin{deluxetable}{lrrrrrrrrrrrrrrrrrrrrrrr}
\tablecolumns{17}
\tabletypesize{\scriptsize}
\tablecaption{Silicon relative to iron abundances based on optical \ion{Si}{1} lines under LTE  and NLTE analyses \label{tbl-7}}
\tablewidth{0pt}
\tablehead{
\colhead{} & \multicolumn{2}{c}{5701 (\AA)}  & \multicolumn{2}{c}{5772 (\AA)} & \multicolumn{2}{c}{6142 (\AA)} & \multicolumn{2}{c}{6145 (\AA)} & \multicolumn{2}{c}{6155 (\AA)} & \multicolumn{2}{c}{6237 (\AA)} & \multicolumn{2}{c}{6243 (\AA)} & \multicolumn{2}{c}{6244 (\AA)}\\
\cline{2-3} \cline{4-5} \cline{6-7} \cline{8-9} \cline{10-11} \cline{12-13} \cline{14-15} \cline{16-17}\\
\colhead{Star} & \colhead{LTE} & \colhead{NLTE} & \colhead{LTE} & \colhead{NLTE} & \colhead{LTE} & \colhead{NLTE}& \colhead{LTE} & \colhead{NLTE} & \colhead{LTE} & \colhead{NLTE} & \colhead{LTE} & \colhead{NLTE} & \colhead{LTE} & \colhead{NLTE} & \colhead{LTE} & \colhead{NLTE}
}
\startdata
Arcturus   &   0.32  &   0.27    &            &            &   0.31    &    0.28   &           &            & 0.38  &  0.28  & 0.36  &  0.30  & 0.36  &  0.32   &       &        \\
HD\,87     &   0.15  &   0.12    &    0.19    &    0.13    &   0.12    &    0.11   &   0.14    &    0.13    & 0.13  &  0.13  &       &        &       &         &       &        \\
HD\,6582   &   0.27  &   0.27    &    0.27    &    0.26    &           &           &   0.27    &    0.27    & 0.26  &  0.25  & 0.23  &  0.23  & 0.31  &  0.31   & 0.27  &  0.27  \\
HD\,6920   &         &           &    0.06    &    0.03    &           &           &   0.00    &  $-$0.01   & 0.10  &  0.05  &       &        &       &         &       &        \\
HD\,22675  &   0.11  &   0.08    &    0.14    &    0.07    &   0.07    &    0.05   &   0.08    &    0.06    & 0.17  &  0.10  &       &        &       &         &       &        \\
HD\,31501  &   0.21  &   0.20    &            &            &   0.18    &    0.18   &   0.23    &    0.22    & 0.22  &  0.20  &       &        & 0.25  &  0.24   & 0.22  &  0.21  \\
HD\,58367  &   0.15  &   0.11    &            &            &   0.11    &    0.11   &   0.13    &    0.13    & 0.23  &  0.15  &       &        &       &         &       &        \\
HD\,67447  &   0.11  &   0.07    &    0.15    &    0.08    &   0.06    &    0.05   &   0.10    &    0.09    & 0.17  &  0.08  & 0.15  & 0.10   & 0.08  &  0.06   & 0.13  &  0.09  \\
HD\,102870 & $-$0.09 & $-$0.10   &  $-$0.05   & $-$0.07    & $-$0.08   & $-$0.08   & $-$0.06   & $-$0.06    &$-$0.03&$-$0.05 &$-$0.09&$-$0.10 &$-$0.07& $-$0.07 &$-$0.09&$-$0.09 \\
HD\,103095 &  0.36   &   0.35    &            &            &           &           &           &            & 0.25  &  0.25  & 0.30  &  0.30  & 0.32  &  0.32   & 0.28  &  0.28  \\
HD\,121370 &         &           &    0.21    &    0.18    &   0.17    &    0.17   &   0.17    &    0.17    & 0.29  &  0.23  & 0.24  &  0.21  &       &         &       &        \\
HD\,148816 &         &           &    0.18    &    0.17    &    0.14   &    0.14   &   0.23    &    0.23    & 0.17  &  0.16  & 0.17  &  0.17  & 0.18  &  0.18   & 0.19  &  0.19  \\
HD\,177249 &  0.04   &   0.02    &            &            &   0.05    &    0.05   &   0.02    &    0.02    & 0.13  &  0.07  &       &        &       &         &       &
\enddata
\end{deluxetable}
\clearpage
\end{landscape}

\clearpage
\begin{landscape}
\begin{deluxetable*}{lrrrrrrrrrrrrr}
\tablecolumns{14}
\tabletypesize{\scriptsize}
\tablecaption{Equivalent widths of neutral iron lines for sample stars.  \label{tbl-8}}
\tablewidth{0pt}
\tablehead{
\colhead{$\lambda$ (\AA)} & \colhead{Arcturus} & \colhead{HD\,87}  & \colhead{HD\,6582}  & \colhead{HD\,6920} & \colhead{HD\,22675} & \colhead{HD\,31501}  & \colhead{HD\,58367} & \colhead{HD\,67447} & \colhead{HD\,102870}  & \colhead{HD\,103095} & \colhead{HD\,121370} & \colhead{HD\,148816}  & \colhead{HD\,177249}
}
\startdata
4661.534 	&	$--$	&	$--$	&	18.0 	&	39.8 	&	$--$	&	$--$	&	$--$	&	89.5 	&	40.4 	&	8.0 	&	48.7 	&	10.4 	&	$--$	\\
4808.149 	&	$--$	&	$--$	&	11.5 	&	25.6 	&	$--$	&	$--$	&	$--$	&	72.4 	&	28.8 	&	8.4 	&	37.1 	&	7.1 	&	$--$	\\
4885.430 	&	$--$	&	$--$	&	55.7 	&	75.3 	&	$--$	&	$--$	&	$--$	&	126.2 	&	77.8 	&	38.1 	&	87.9 	&	37.7 	&	$--$	\\
5223.186 	&	64.3 	&	52.0 	&	14.7 	&	32.5 	&	61.5 	&	$--$	&	$--$	&	67.8 	&	30.7 	&	8.7 	&	37.3 	&	7.2 	&	$--$	\\
5242.497 	&	117.5 	&	115.9 	&	63.6 	&	92.2 	&	120.9 	&	$--$	&	146.5 	&	146.4 	&	92.0 	&	54.9 	&	108.0 	&	54.8 	&	121.6 	\\
5379.579 	&	90.9 	&	87.5 	&	38.1 	&	59.5 	&	93.8 	&	$--$	&	111.4 	&	114.7 	&	63.7 	&	22.4 	&	76.1 	&	27.7 	&	93.4 	\\
5398.279 	&	91.1 	&	94.7 	&	48.3 	&	71.0 	&	104.2 	&	$--$	&	116.8 	&	124.5 	&	78.9 	&	33.2 	&	88.7 	&	38.5 	&	102.9 	\\
5522.449 	&	61.0 	&	66.2 	&	20.1 	&	41.5 	&	71.1 	&	$--$	&	78.8 	&	84.3 	&	44.1 	&	10.6 	&	57.0 	&	14.3 	&	69.8 	\\
5546.506 	&	72.2 	&	75.1 	&	26.0 	&	51.0 	&	81.7 	&	$--$	&	91.6 	&	100.4 	&	53.4 	&	14.1 	&	68.8 	&	17.7 	&	80.4 	\\
5618.633 	&	71.0 	&	69.4 	&	27.4 	&	50.3 	&	76.0 	&	42.3 	&	88.9 	&	93.6 	&	53.2 	&	12.3 	&	65.3 	&	18.4 	&	77.6 	\\
5651.469 	&	32.2 	&	33.5 	&	6.6 	&	18.8 	&	39.7 	&	14.6 	&	45.7 	&	45.1 	&	19.2 	&	$--$	&	27.3 	&	$--$ 	&	35.8 	\\
5679.023 	&	70.0 	&	76.4 	&	32.8 	&	61.2 	&	81.9 	&	53.7 	&	87.9 	&	97.2 	&	60.8 	&	17.1 	&	72.0 	&	24.0 	&	81.5 	\\
5793.915 	&	53.6 	&	58.6 	&	12.9 	&	29.8 	&	62.1 	&	28.1 	&	70.0 	&	84.6 	&	34.4 	&	6.1 	&	45.9 	&	9.3 	&	72.0 	\\
5853.148 	&	$--$	&	$--$	&	$--$	&	$--$	&	44.2 	&	12.0 	&	45.1 	&	51.5 	&	6.0 	&	$--$ 	&	10.5 	&	$--$ 	&	27.4 	\\
5855.077 	&	34.2 	&	39.9 	&	7.8 	&	22.0 	&	43.9 	&	17.5 	&	$--$	&	49.5 	&	23.0 	&	$--$ 	&	31.2 	&	5.3 	&	39.2 	\\
5929.677 	&	54.7 	&	63.3 	&	16.3 	&	$--$	&	67.6 	&	37.7 	&	73.9 	&	78.0 	&	40.4 	&	17.0 	&	50.9 	&	17.0 	&	63.4 	\\
6024.058 	&	116.8 	&	125.7 	&	87.0 	&	107.5 	&	132.6 	&	120.1 	&	$--$	&	156.0 	&	114.0 	&	70.3 	&	124.7 	&	66.7 	&	135.0 	\\
6078.491 	&	81.2 	&	93.8 	&	47.5 	&	76.6 	&	99.0 	&	81.2 	&	$--$	&	117.0 	&	82.5 	&	29.5 	&	92.8 	&	35.2 	&	99.2 	\\
6079.009 	&	58.3 	&	62.9 	&	19.3 	&	40.7 	&	69.7 	&	41.6 	&	80.0 	&	85.4 	&	48.9 	&	9.2 	&	54.2 	&	13.7 	&	70.1 	\\
6151.623 	&	114.6 	&	88.1 	&	31.3 	&	47.5 	&	97.9 	&	51.4 	&	117.4 	&	122.1 	&	45.4 	&	21.4 	&	54.7 	&	17.4 	&	91.5 	\\
6173.335 	&	133.0 	&	109.5 	&	48.3 	&	70.3 	&	119.9 	&	52.3 	&	146.8 	&	148.9 	&	66.0 	&	40.0 	&	78.0 	&	33.7 	&	115.1 	\\
6200.321 	&	132.7 	&	111.0 	&	52.2 	&	73.0 	&	121.7 	&	74.2 	&	140.3 	&	148.0 	&	78.8 	&	41.4 	&	87.1 	&	34.5 	&	116.4 	\\
6240.646 	&	114.7 	&	$--$	&	29.8 	&	43.7 	&	$--$	&	52.3 	&	$--$	&	127.4 	&	43.2 	&	19.8 	&	55.2 	&	16.5 	&	$--$	\\
6322.686 	&	140.7 	&	$--$	&	53.8 	&	76.8 	&	$--$	&	76.7 	&	$--$	&	150.6 	&	73.0 	&	44.3 	&	91.0 	&	40.6 	&	$--$	\\
6335.331 	&	132.6 	&	$--$	&	85.1 	&	103.5 	&	$--$	&	109.1 	&	$--$	&	196.0 	&	98.4 	&	79.2 	&	115.3 	&	66.2 	&	$--$	\\
6481.877 	&	$--$	&	$--$	&	$--$	&	63.9 	&	$--$	&	72.4 	&	$--$	&	142.9 	&	62.4 	&	34.1 	&	76.6 	&	27.5 	&	$--$	\\
6593.871 	&	$--$	&	$--$	&	68.8 	&	89.1 	&	$--$	&	90.6 	&	$--$	&	177.7 	&	93.9 	&	61.3 	&	101.1 	&	55.4 	&	$--$	\\
6726.666 	&	$--$	&	$--$	&	19.8 	&	50.0 	&	$--$	&	42.1 	&	$--$	&	86.6 	&	49.6 	&	9.8 	&	62.8 	&	15.3 	&	$--$	\\
6839.831 	&	$--$	&	$--$	&	15.3 	&	$--$	&	$--$	&	32.2 	&	$--$	&	94.5 	&	28.9 	&	8.2 	&	43.5 	&	7.2 	&	$--$	\\
6857.250 	&	$--$	&	$--$	&	$--$	&	$--$	&	$--$	&	18.8 	&	$--$	&	55.3 	&	$--$	&	$--$ 	&	$--$	&	5.9 	&	$--$	
\enddata
\tablecomments{Seriously blending lines and the lines with bad S/N are rejected during determining the stellar parameters for a given star.}
\end{deluxetable*}
\clearpage
\end{landscape}

\end{document}